\documentclass[a4paper,11pt]{article}
\pdfoutput=1  
\usepackage{jheppub}  
\usepackage{mathtools}
  
 \def\antinor{{^{_\circ}_{^\circ} }}
 \def\Zh{Zhukovsky\ } \makeatother   
 \def\defeq{\stackrel{\text{def}}=}
 \newcommand\re[1]{({\ref{#1}})}
 \def\be{\begin{eqnarray} } \def\ee{\end{eqnarray}} \def\z{\zeta}
 \def\IR{{\mathbb{R}}} 
\def\IZ{{\mathbb{Z}}} 
\def\IK{{\mathbb{K}}}

 \def\C {{\mathbb{C}}} 
  
\def\IO{{\mathbb{O}}}

\def\psu{{\mathfrak{psu} }}

\def\UU{\mathbf{U}} 
 \def\no{\nonumber} \def\la{\label} 
     \def\even{ \mathrm{e} }\def\odd{ \mathrm{o} }
    
     \def\({\left(} \def\){\right)} \def\<{\langle\,} \def\>{\,
   \rangle} \def\[{\left[} \def\]{\right]} \def\tr{{\rm Tr} }
     \def\hf{ {\textstyle{1\over 2}} } 
      \def\CK{{ \mathcal{ K} }}
 \def\CA{{\cal A}}
   
\def\CO{{ \mathcal{ O} }}
 \def\CH{{ \mathcal{ H} }} 
    \def\CN{{ \cal N}} 
     \def\a{\alpha}
     \def\vp{\varphi} 
      \def\s{\sigma} \def\t{\tau} \def\th{\theta}
      \def\p{\partial}
      
       \def\G{\Gamma} 
      \def\b{\beta}   
 \def\d{\delta}
\def\W{ W}
 \def\JJ{\mathbf{J}}
 \def\RR{\mathbf{R}}
    \def\C{\mathbf{C}}

    \def\K{ \mathbf{K}}
     
   \def\NN{ \mathbf{N}}
    
\def\pf{\mathrm{Pf}}
 \def\MM{ \mathbf{M}} 
  \def\XX{ \mathbf{X}} 
   \def\CKI{ \overset{_\circ}{ \CK }}
   \def\KI{ \overset{_\circ}{ \mathbf{K}}  }
  \def\QQ{ {\mathbf{Q}} } 
  \def\OO{\mathbf{\Omega}} 
   \def\SS { \mathbf{S}}
     \def\I{\mathbf{I}}
\def\X{\mathrm{X}} 
 \def\PP{\mathbf{P}} 
 \def\gell{  {_{\ge\ell}}}
 \def\lell{ {_{<\ell}}}

\DeclareMathAlphabet{\mathbbmsl}{U}{bbm}{m}{sl}
\def\slK{\mathbbmsl{K}}

\title{\boldmath Octagon with finite bridge:  
free fermions and determinant identities }

 \author[a]{Ivan Kostov\note{Corresponding author.}}
 \author[b]{ and Valentina B. Petkova}

\affiliation[a]{Universit\'e Paris-Saclay, CNRS, CEA, Institut de
physique th\'eorique, 91191, Gif-sur-Yvette, France}
\affiliation[b]{Institute for Nuclear Research and Nuclear Energy,
Bulgarian Academy of Sciences, Sofia, Bulgaria}

\emailAdd{ ivan.kostov@ipht.fr }
\emailAdd{vbpetkova@yahoo.com}

  \abstract{ We continue the study of the octagon form factor which
  helps to evaluate a class of four-point correlation functions in
  $\CN=4$ SYM theory.  The octagon is characterised, besides the
  kinematical parameters, by a ``bridge'' of $\ell$ propagators
  connecting two non-adjacent operators.  In this paper we construct
  an operator representation of the octagon with finite bridge as an
  expectation value in the Fock space of free complex fermions.  The
  bridge $\ell$ appears as the level of filling of the Dirac sea.  We
  obtain determinant identities relating octagons with different
  bridges, which we derive from the expression of the octagon in terms
  of discrete fermionic oscillators.  The derivation is based on the
  existence of a previously conjectured similarity transformation,
  which we find here explicitly.
}

\begin{document} 
\maketitle
\flushbottom

\section{Introduction}
\label{sec:intro}

 A new non-perturbative approach for the computation of the
 correlation functions of single-trace operators in the $\CN=4$
 supersymmetric Yang-Mills theory has been developed in the past five
 years \cite{BKV1,Fleury:2016ykk,Eden:2016xvg, Fleury:2017eph,
 Bargheer:2017nne, Bargheer:2018jvq}.  The method, mostly referred to as
 hexagonalisation, is based on the world-sheet integrability of
 $\CN=4$ SYM \cite{Minahan:2002ve}.  The hexagonalisation prescribes
 to decompose the correlation function into elementary blocks called
 hexagon form factors, or shortly hexagons, which are almost uniquely
 determined by the huge symmetry of the theory.  Being formulated in
 terms if infinite-volume form factors, the prescription involves
 divergencies and, in spite of some important progress \cite{BGK}, it
 still awaits an appropriate regularisation procedure.

Remarkably, a class of four-point functions of half-BPS operators with
large R-charges and specially tuned polarisations, discovered in
\cite{Coronado:2018ypq, Coronado:2018cxj}, are free of divergencies
and can be evaluated exactly for any value of the 't Hooft coupling.
In these correlation functions the hexagons couple only pairwise.  The
composite form factors representing two paired hexagons, named {\it
octagons}, completely factorise.  The factorisation was shown to take
place in all orders of the $1/N_c$ expansion \cite{Bargheer:2019kxb}.
If there are $\ell$ propagators sandwiched between the two hexagons,
one speaks of octagon with {\it bridge} $\ell$.

The two constituent hexagons are bound by exchanging virtual particles
in the mirror channel.  In \cite{Kostov:2019stn,Kostov:2019auq}, the
octagon was represented as a Fredholm pfaffian and was also given a
more tractable representation as the pfaffian of a discrete kernel
$\K$ representing a complex semi-infinite anti-symmetric matrix
$\K_{m,n}$ with $m,n\ge 0$.  It was also conjectured that the octagon
kernel can be rotated to a simplified kernel $\KI$ which is a real
half-sparse matrix and as such can be split into two equivalent
diagonal blocks.  Based on this conjecture, the pfaffian was expressed
as the determinant of one of the blocks.  The simplified kernel $ \,
\KI $ was defined in \cite{Kostov:2019auq} by its perturbative series
and then non-perturbatively in
\cite{Belitsky:2019fan,Belitsky:2020qrm, Belitsky:2020qir}.  This
second representation of the octagon allowed the authors of
\cite{Belitsky:2019fan,Belitsky:2020qrm, Belitsky:2020qir} to
reformulate the latter as Fredholm determinant of a generalised Bessel
kernel, for which powerful mathematical methods have been developed
previously.  However the existence of a similarity transformation
turning $\K$ into $\KI$ has not been established.  One of the goals of
this paper is to construct explicitly such a
transformation.\footnote{While we were working on this manuscript, we
learned that Andrey Belitsky and Gregory Korchemsky found another
solution for the similarity transformation, to be published as
appendix to v2 of \cite{ Belitsky:2020qir}.  We comment on their
solution in our appendix \ref{Appendix:B}.  The two solutions are
related by a transformation which leaves the kernel $\K$ invariant.  }

The octagon is the simplest of a family of computable observables in
$\CN=4$ SYM, such as the cusp anomalous dimension \cite{BES} and the
MHV six-gluon amplitude in the collinear \cite{Basso:2015uxa} and
close-to-the origin \cite{Basso:2020xts} limits.  As emphasised in
\cite{Basso:2020xts}, these objects exhibit similar mathematical
structures involving semi-infinite matrices.

In this paper we propose an operator description for the octagon based
on a pair of complex fermionic fields, $\psi(x)$ and $\psi^*(x)$, with
the holomorphic variable $x$ being the \Zh parametrisation of the
rapidities of the virtual particles.  Similar descriptions exist for
all observables mentioned above.  Below we present, for reader's
convenience, a short summary of our main results.

The operator formalism proposed here is a Fock space realisation of
the description with real fermions presented in \cite{Kostov:2019auq}.
The Fock space for the complex fermions is a direct sum of sectors
characterised by the $U(1)$ charge of the vacuum or, in other words,
by the level of filling of the Dirac sea.  The octagon with bridge
$\ell$ is constructed as the expectation value of a product of
exponential operators in the sector of charge $\ell$,
   \be
   \begin{aligned}
\la{operoct} \IO_\ell & = \<\ell| \exp[\hf \psi\K\psi] \ \exp[-\hf
\psi^ * \C \psi^ * ] |\ell\> \,,
   \end{aligned}
   \ee
$\K$ is the octagon kernel and $\C$ is a standard quasi-diagonal
symplectic matrix.  The right exponential imposes non-trivial
correlation of the modes of $\psi$ and resembles the operators of
boundary states in CFT, hence the notation
 \be
 \la{defchovchica}
  |\,\ell \>\!\!\>\defeq   \exp[-\hf  \psi^ * \C \psi^ * ] |\ell\> .
  \ee
As the two exponents contain either two creation operators, or two
annihilation operators, the $U(1)$ charge is not preserved and the
expectation value is given by a Fredholm pfaffian
\cite{Kostov:2019auq}.

The Fock-space realisation \re{operoct} gives a nice interpretation of
the bridge as an operator composed of the $\ell$ lowest fermion
oscillator modes.  Based on this we show that the octagon with
non-zero bridge $\ell$ is obtained by multiplying the octagon with
$\ell=0$ by a pfaffian of a $2\ell\times 2\ell $ matrix of fermionic
correlators.

We give an explicit solution for the similarity transformation
mentioned above and explore its consequences for the fermionic
oscillator model.  For any $\ell\ge 0$, the similarity transformation
acts only on the oscillators $\psi_n,\psi^*_n$ above the Fermi level,
$n\ge \ell$, by a semi-infinite matrix $\UU_\ell$,
\be \la{changeofbas} \tilde \psi _j =\sum_{k\ge \ell} [\UU_\ell ]_{jk}
\ \psi _ k , \ \ \tilde{ \psi^ * \!\!  } _{ j} = \sum_{k\ge
\ell}[\UU^{-1}]_{kj} \psi ^*_k\,,  \ \  \  j\ge \ell \,.\ee
The canonical transformation \re{changeofbas} preserves the matrix
$\C$ and transforms the octagon kernel $\K$ into the simplified kernel
$\KI$.  We will give its explicit formula for any $ \ell$, but what is
important is the very fact of its existence.  The operator expression
for the octagon then takes the form
  \be\begin{aligned} \la{operoctb} \IO_\ell &= \<\ell| \exp[\hf \psi
  \KI \psi ] \ \exp[-\hf \psi ^*\C \psi ^*] |\ell\>.  \end{aligned}\ee
Here we replaced $\{\tilde \psi, \tilde \psi^*\}$ by $\{\psi,
\psi^*\}$, as the existence of a transformation \re{changeofbas} for
any $\ell$ then guarantees that the vacuum states have the same form
for the original and the transformed fermions.

Both the simplified kernel $\KI$ and the matrix $\C$ relate only modes
of different parity.  Thanks to this property, half of the modes in
\re{operoctb} can be eliminated and the resulting operator expression
is exponential of a fermion bilinear which, unlike the exponential
operators in \re{operoctb}, preserves the $U(1)$ charge.  This leads
to the Fredholm determinant formula for the octagon and to finite
determinant relations between octagons with different bridges.
 
For even/odd bridge we expressed the octagon as an expectation value
in the Fock space built on the odd/even oscillators, $\psi^\even _j=
\psi_{2j}, \psi^{*\even}_j = \psi^*_{2j}$ and $\psi^{\odd}_j=
\psi_{2j+1}, \psi^{*\odd}=\psi^*_{2j+1}$,
 \be \la{improveddet} 
  \IO_{\ell } =
  \begin{cases}
     \< m,\odd | \ e^{- \psi^\odd
  \IK^{\odd\odd} \psi^{*\odd} } |m,\odd\>\, = \det [\( 1-
  \IK^{\odd\odd}\)_{\ge m}]\,  & \text{ if} \ \ell = 2m, \\
  \< m,\even | e^{ - \psi^\even \, \IK^{\even \even } \psi^{*\even } }
  |m,\even \>\, = \det [\( 1- \IK^{\even\even}\)_{\ge m }] & \text{
  if} \ \ell = 2m-1.
\end{cases}
   \ee
The vacuum states $|m,\odd\>$ and $ |m,\even \>\,$ in \re{improveddet}
are the standard vacuum vectors of charge $m$ respectively for the
ensembles of odd and the even oscillator modes.  By $
\IK^{\even\even}$ and $ \IK^{\odd\odd}$ we denoted respectively the
even-even and the odd-odd blocks of the block diagonal product $\KI\C=
\IK^{\even\even}\oplus \IK^{\odd\odd}$ and $1$ stays for the identity
matrix.  Finally, for any semi-infinite matrix $\mathbf{A} =\{
\mathbf{A}_{i,j}\}_{i,j\ge 0}$, the symbol $(\mathbf{A})_{\ge m}$
denotes the semi-infinite matrix obtained by deleting the first $m$
rowes and columns, $ (\mathbf{A})_{\ge m} =\{
\mathbf{A}_{i,j}\}_{i,j\ge m} $.  The determinants in \re{improveddet}
are equivalent to those formulated in
\cite{Kostov:2019auq,Belitsky:2019fan}, only the matrix elements are
indexed differently.
  
The operator representations in the form \re{improveddet} give rise to
$m\times m$ determinant identities, presented in section
\ref{section:dets}, which relate the octagons with finite bridge $\ell
= 2m-1$ or $\ell= 2m$ to the octagon with zero bridge.  Hence the
ratio $\IO_{2m}$ and $\IO_0$ as an $m\times m$ determinant,
\be \la{Reedet} {\IO_{2m}\over \IO_0} =\det\[ (1+ \IR
^{\odd\odd})_{<m}\]
,\qquad {\IO_{2m-1}\over \IO_0} =   \det\[\( 1+\IR ^{\even\even}
 \)_{< m}\],
\ee
where $\IR^{\a\a}$ are the even ($\a=\even$) and odd ($\a=\odd$)
resolvent matrices
   \be \la{defresolv} \IR ^{\even\even}= {\IK^{\even\even}\over 1-
   \IK^{\even\even}}, \qquad \IR ^{\odd\odd} = {\IK^{\odd\odd}\over 1-
   \IK^{\odd\odd}} , \ee
and the symbol $(\mathbf{A})_{< m}$ denotes the $m\times m$ diagonal
block $\{ \mathbf{A}_{i,j}\}_{0\le i,j\le m-1} $ of the semi-infinite
matrix $ \mathbf{A}$.  For example,
 \be \la{exevenbr} \begin{aligned}
{\IO_2\over \IO_0} &= 1+ \IR ^{\odd\odd}_{0,0} ,
\qquad {\IO_{1}\over \IO_0}  =1+\IR ^{\even\even}_{0,0},\\
 {\IO_{3}\over \IO_0} & =(1+\IR ^{\even\even}_{0,0})(1+\IR
 ^{\even\even}_{1,1})-\IR ^{\even\even}_{0,1}\IR ^{\even\even}_{1,0},
 \\
 {\IO_4\over \IO_0} &= (1+ \IR ^{\odd\odd}_{0,0})(1+ \IR
 ^{\odd\odd}_{1,1}) - \IR ^{\odd\odd}_{0,1} \IR ^{\odd\odd}_{1,0} .
\end{aligned}
\ee

The organisation of the paper is as follows.  In section \ref{sect:2}
we derive, starting from the expression of the octagon as a sum over
virtual particles, the operator representation in terms of fermion
oscillators.  From the fermionic representation we re-derive the
expression for the octagon as semi-infinite pfaffian found in
\cite{Kostov:2019auq} as well as new finite pfaffian formulas relating
octagons with different bridges.  In Section
\ref{section:similaritytr} we give an explicit expression for the
similarity transformation $ \UU_\ell $ relating the original and the
improved octagon kernels for any $\ell$.  The details of the proof are
relegated to appendices \ref{Appendix:A} and \ref{Appendix:C}.  For
$\ell=0$, we give an alternative exponential expression for the
similarity transformation, the derivation of which is presented in
appendix \ref{Appendix:D}.  In section \ref{section:dets} we derive
the operator representations \re{improveddet} and the finite
determinant formulas that follow from them.  Section
\re{sect:conclusions} contains some comments on the results.

\section{The octagon from free fermions}
\la{sect:2}

 \subsection{The sum over virtual particles as a Coulomb gas }
 \la{ss:Coulombgas}

The role of this subsection is to remind the notations and make the
presentation self-consistent.  The octagon $\IO_\ell = \IO_\ell(z,
\bar z, \a, \bar\a)$ is characterised by four points $x_1,...,x_4$ in
the Euclidean space and four polarisations $y_1,..., y_4$, as well as
by the length $\ell$ of the bridge separating the two hexagons which
should be crossed by the virtual particles.  The bridge summarises the
effect of a stack of $\ell$ tree-level propagators connecting the
operators $\CO_1$ and $\CO_4$.  The octagon is also a function of the
't Hooft coupling $g$.  The trivial dependence of the (large)
$R$-charges of the four half-BPS operators is factored out.  Thanks to
the conformal symmetry, the dependence on $x_i, y_i$ is only through
the cross ratios in the coordinate and in the flavour spaces
  \be
  \begin{aligned}
  \la{crossratios} z\bar z &= {x_{12}^2 x_{34}^2\over x_{13}^2
  x_{24}^2} ,\ \qquad (1-z)(1-\bar z)= {x_{14}^2 x_{23}^2\over
  x_{13}^2 x_{24}^2} , \\
  \alpha\bar\alpha &= { y_{12} ^2 y_{34}^2 \over y_{13}^2 y_{24}^2} ,
  \ \qquad (1-\alpha)(1-\bar\alpha) = {y_{14}^2 y_{23}^2\over y_{13}^2
  y_{24}^2}.
    \end{aligned}
  \ee
where $x_{ij}^2= (x_i-x_j)^2, y_{ij}^2 = (y_i- y_j)^2$ and $y_i^2=0$.
For the cross ratios in the Euclidean space we adopt the exponential
parametrisation
\be \la{angles} \begin{aligned} z &= {e^{-\xi+i\phi } } , \quad \bar z
= {e^{-\xi - i\phi} } , \quad \a = e^{\vp - \xi + i\th}, \quad \bar\a
= e^{\vp -\xi - i \th}.  \end{aligned}\ee
The parameters $\phi$ and $\xi$, respectively $\vp$ and $\th$,
characterise the rotation aligning the two hexagons in the Euclidean,
respectively flavour, space.  We consider Euclidean metric where the
angle $\phi$ is real.  In Minkowski space the angle $\phi$ should be
taken complex, $\phi = \pi + i y$ with $y$ real.

The octagon represents two hexagons glued together by inserting a
complete set of virtual states in the Hilbert space associated with
the common mirror edge.  An $n$-particle virtual state is
characterised by the rapidities $u_i$ and the bound-state numbers
$a_i$ of its particles.  The contribution of such virtual state
factorises into one-particle factors $W_{a_j}(u_j)$ and two-particle
interactions $W_{a_j, a_k}(u_j,u_k)$ accounting for the hexagon
weights.  The octagon thus is expanded as a series of multiple
integrals with integrand given by a product of local and bi-local
weights \cite{Coronado:2018ypq}
 \be
 \begin{aligned}
  \la{defIn} \mathbb{\IO}_{\ell}&= \hf \sum_{\pm} \sum_{n=0}^\infty {
  1\over n!} \sum_{a_1,..., a_n\ge 1} \int \prod_{j=1}^n {du_ j \over
  2\pi i }\ \W^\pm _{a_j}(u_j) \ \prod_{j<k}^n \W_{a_j, a_k}
  (u_j,u_k).
  \end{aligned}
 \ee

\noindent {\bf \it $\bullet$ Bi-local weights.  } \ \ The bi-local
weights are defined in terms of a single function
\be \la{defWu} \W (u,v)= {x(u)-x(v)\over x(u)x(v)-1} \ee
  where the function $x(u)$ is defined by the \Zh map
  \begin{align}\la{paramux}
{u/g} = x+{1/x}
\end{align}
transforming the physical sheet in the rapidity plane into the
exterior of the unit circle.  Namely
\be \W_{a,b}(u,v) = \W(u\!  +\!  \textstyle{i\over 2}a, v\!  +\!
\textstyle{i\over 2}b) \, \W(u\!  +\!  \textstyle{i\over 2}a, v\!  -
\!  \textstyle{i\over 2}b) \, \W(u\!  - \!  \textstyle{i\over 2}a, v\!
+\!  \textstyle{i\over 2}b) \, \W(u\!  - \!  \textstyle{i\over 2}a,
v\!  - \!  \textstyle{i\over 2}b) .  \ee
 {\bf \it $\bullet$ Local weights.  } \ \ The one-particle factors are
  \be \la{measuredma} \W^\pm _a(u,\xi) = {1\over g } (-1)^a \chi^\pm
  _a \ \,\Omega_\ell(u + \textstyle{i\over 2}a , \xi) \ \Omega_\ell(u
  - \textstyle{i\over 2}a, \xi) \, \times \W(u+ \textstyle{i\over 2}a,
  u- \textstyle{i\over 2}a).  \ee
where 
\be \la{defOm} \Omega _\ell(u,\xi) = {1\over x^{\ell }}\, {e^{ig \xi\,
[x-1/x]} \over x-1/x} = g \, {e^{ig \xi\, [x-1/x]} \over x^\ell} \
{d\log x\over du} \, , \ee
 $\chi^\pm _a$ is essentially the character of the $a$-th
 antisymmetric representation of $\psu(2|2)$
\be\begin{aligned} \la{su22charakiri} \chi^\pm _a (\phi, \vp, \th ) =
(-1)^a \ {\sin (a\phi )\over \sin\phi }\ \left[ 2\cos \phi -2\cosh
(\vp \pm i\th) \right] .  \end{aligned}
\ee
For simplicity we will assume that $\th=0$.  The function $\Omega_\ell
(u,\xi)$ reflects the form of the momentum and the energy of the
mirror magnons as functions of the rapidity $u$,
   \be\begin{aligned} \tilde p_a(u) &= \textstyle{1\over 2} g
   (x-{1\over x})_{u+ia/2}+ \hf g(x-{1\over x})_{u-ia/2}, \\
\quad  \tilde E_a(u) &=   \log x|_{u+ia/2}+   \log x|_{u-ia/2} .
\end{aligned}
\ee
The $\psu(2|2)$ characters are determined by the generating function
 \be\begin{aligned}\la{diffoTa} \mathcal{W}  (t) & =
1+  \sum_{a=1}^\infty (-1)^a \chi  _a \, e^{-a t}
 = 1- {\cosh \vp -\cos\phi\over \cosh t - \cos\phi} .
\end{aligned}
 \ee

 \subsection{ Free  complex fermions}
\la{ss:Fockff}

The fermionic representation we give here was sketched in
\cite{Li-2019}.  Let us first give our conventions, mostly following
the conventions of \cite{JimboMiwa-tau}, with $\psi_{\mathrm{here}}=
\psi^*_{\mathrm{there}}$, $\psi^*_{\mathrm{here}}=
\psi_{\mathrm{there}}$.  The pair of fermionic fields is defined as
\be
\begin{aligned}\la{fermionx}
\psi(x)&= \sum_{n\in \mathbb{Z} }\psi_{n } x^{ -n } , \quad \psi
^*(x)= \sum_{n\in \mathbb{Z} } \psi^ * _{n} \ x ^{ n }, \\
& [\psi_m,   \psi^ * _n]_+ =\delta_{m,n}, \qquad m,n\in \IZ.
\end{aligned}
\ee
The operators $\psi_n, \psi^ * _n$ act in the standard fermionic Fock
space $\CH$, which splits as a sum of Fock spaces with given $U(1)$
charge $\ell $,
\be \CH=\underset{\ell \in \IZ} \oplus \CH_\ell .  \ee
The Fock space $\CH_\ell$ is built on the highest-weight state $|\ell
\>$ and its dual $\langle \ell | $, constructed for $\ell\ge 0$ as
\be \la{defvac} \langle \ell | = \langle 0| \prod_{n=0} ^{ \ell -1 }
\psi_n, \quad | \ell \rangle = \prod_{n=0}^{\ell -1 } \psi^ * _n|
0\rangle.  \ee
 The two vacua satisfy
\be
\la{fermvac}
\begin{aligned}
\psi^ * _n | \ell \rangle &= 0, \ \ \langle \ell | \psi_{n} =0 \qquad
(n<\ell ), \\
\langle \ell | \psi^ * _{n} &= 0, \ \psi_n | \ell \rangle = 0\qquad \
(n\ge \ell ) .  \end{aligned}\ee
 The non-vanishing correlators are
\be \la{ccrf} \langle \ell | \psi_m \psi^ * _n|\ell \rangle =
\begin{cases}
    \delta_{m,n} & \text{if } m\ge \ell , \\
   0 & \text{if } m <\ell  
\end{cases}
\ee
 and the two-point function is
  \be \la{propagator} G(x,y) \equiv \langle \ell | \psi(x)\ \psi^ *
  (y)|\ell \rangle _{_{ |y|<|x|}} = { \({y/x}\)^{\ell } \over 1 - y/x}
  .  \ee
The correlation function of a product of fermions is given by the
determinant of the two-point correlators.
 
 In \cite{Kostov:2019auq}, the bi-local weights in the expansion
 \re{defIn} of section \ref{sec:intro} were expressed in terms of the
 two-point function of the field $\psi(x)$ whose form was postulated.
 On the present interpretation the two-point function of the field
 $\psi$ results from replacing the right vacuum by a coherent
 state\footnote{ I.K. is obliged to Y. Matsuo for a discussion on this
 way to introduce fermions.  }

 \begin{align}
 \la{defCC} |\,\ell \>\!\!\> \defeq e^{ - {1\over 2} \psi^ * \C \psi^
 * } | \ell \>\, , \quad \psi ^*\C \psi^ * = \sum_{m,n \ge 0 } \psi^ *
 _{m } \C_{mn} \psi^ * _{n} \,
\end{align}
where $\C$ is the skew-symmetric matrix with elements
\be \la{defCCC} \C_{m,n} = \delta_{m+1, n}- \delta_{ m , n+1}
\qquad (m,n\ge 0).  \ee
For the action of the fermionic oscillators $\psi_n$ on the coherent
state one obtains
\be\begin{aligned} \la{propagatorF} \( \psi_m +[\C\psi ^*]_m \)
|\,\ell \>\!\!\> &=0 \,, \quad \quad m\ge \ell\, .
   \end{aligned} 
   \ee
With the ket vacuum replaced by the coherent state, the
$\psi$-oscillators have a non-vanishing correlation
    \be \< \ell| \psi_m\psi_n |\,\ell \>\!\!\> = \C_{m,n} \ee
and their two-point function takes the desired form in the
$x$-representation \footnote{Eq.  \re{defCC} gives a fermionic
operator realisation of the twisted vertex operators introduced in
\cite{Kostov:2019auq},
$$
  \langle \ell | \ \psi(x)\psi(y) |\,\ell \>\!\!\> =  {1\over (xy)^\ell}{x-y\over xy-1}=
  \langle 0 | :e^{\phi(x)}::e^{\phi(y)}: e^{-{\ell\over \sqrt{2} }
  \hat q}| 0 \rangle.
$$
The rhs represents an expectation on the bosonic vacuum, with $\phi(x)
= {1\over \sqrt{2}}(\vp(x)- \vp(1/x)) $, where $\vp(x)$ being the
standard bosonic oscillator with mode expansion $\vp(x) = \hat q + p
\log x - \sum_{n\ne 0} {J_n\over n} x^{-n}$ with $[J_n,J_m]= n
\d_{n+m,0}, \ [\hat p,\hat q]=1$, and the action of the bosonic
oscillators on the bosonic vacua is $\<0| J_{n<0}=\<0| \hat q=
J_{>0}|0\> = \hat p|0\> =0 $.  }
  \be \la{defCxbisbis} \langle \ell | \ \psi(x)\psi(y) |\,\ell
  \>\!\!\> = \sum _{m,n\ge \ell} \C_{mn}\ x^{-m } y^{-n }= (x
  y)^{-\ell} {x-y\over xy-1} .  
  \ee
As in any ensemble of fermions, the $2n$-point correlator is the
pfaffian of the matrix of the two-point correlators:
\be\la{cauchipf} \<\ell \ | \psi(x_1) ...  \psi(x_{2n})|\,\ell
\>\!\!\> = \mathrm{Pf}\left( \left[ {1\over (x_jx_k) ^\ell } {x_j-
x_k\over x_jx_k -1}\right] _{i,j=1}^{2n}\right) = \prod_{i=1}^{2n}
{1\over {x_i ^{ \ell}}} \prod_{j<k}^{2n} {x _j-x_k\over x_jx_k-1}.
\ee

Applying \re{cauchipf}, we can sum up the expansion \re{defIn}.  For
that we take the fermion in the rapidity plane by replacing $x\to
x(u)$ in the expansion \re{fermionx}.  A virtual particle of type $a$
is represented by the fermion pair $\psi[x(u+ i a/2)]\psi[x(u- i
a/2)]$.  Its expectation value yields the last factor in local weights
\re{measuredma}.  All the bi-local weights are nicely reproduced by
the correlation functions of these fermion pairs and the expansion
takes the form
 \be
 \begin{aligned}
  \la{defIn2} \mathbb{\IO}_{\ell}&= \sum_{n=0}^\infty { g^{-n}\over
  n!} \!  \sum_{a_1,..., a_n\ge 1} \int \prod_{j=1}^n {du_ j \over
  2\pi i }\ (-1)^{a_j}\chi_{a_j} \ \Omega_0(u_j+\hf i{a_j }, \xi) \
  \Omega_0(u_j- \hf i{a_j },\xi) \\
 &\times \<\ell \ |\prod_{j=1}^n \psi[x(u_j+\hf i a_j)] \
 \psi[x(u_j-\hf i a_j) ]|\ \ell \>\!\!\>.  \end{aligned}
 \ee
The series \re{defIn2} sums up into an exponential,
  \be\begin{aligned} \la{operoctal} \mathbb{\IO}_{\ell} &= \<\ell \ |
  e^{ {1\over 2} \psi K\psi}| \,\ell \>\!\!\> , \\
   \psi K\psi &= {2\over g} \sum_{a\ge 1} (-1)^a \chi_a
(\phi, \vp, \th ) \int _\IR{d u\over 2\pi i} [\Omega _0 \psi ]_{u+
ia/2} [\Omega_0\psi ]_{u- ia/2} .
  \  \end{aligned}\ee
By Fourier transformation the summation in $a$ is separated and gives
the generating function \re{diffoTa} as a function of the Fourier
variable $t$.  The Fourier transforms of the two factors in
\re{operoctal} are given by integrals over real variables $u$ and $v$
running below and above the real axis respectively.  They are
transformed into contour integrals in Zhukovsky variables $x(u)$ and
$y(v)$ which can be deformed to integrals on the unit circle imposing
a bound from below on the $t$-integration
 \be \begin{aligned} \mathbb{\IO}_{\ell} &= \<\ell \ | \exp \( {1\over
 2} {1\over (2\pi i)^2} \oint {d x\over x } \oint {d y\over y } \
 \psi(x) K(x, y) \psi(y) \) | \ell \>\!\!\> \ , \\
  \la{defKxy} K(x,y)&=2 \ e^{i g\xi \, (x -{1\over x} +y- {1\over y})}
  {g } \int_{ | \xi |} ^\infty {d t } \sin\[{\textstyle gt(x+{1\over
  x}-y-{1\over y} )}\] \ \, \X(t)\, , \ \ \ \ \ \\
   X( t) &= { \cos \phi - \cosh\xi \over \cos\phi - \cosh { t} }\, .
    \end{aligned} 
    \ee
 In terms of the fermionic oscillators the quadratic form is
 represented by the semi-infinite matrix $\K =\{ \K_{m,n}\}_{m,n\ge
 0}$.  Using the integration formula
 \be {1\over 2\pi i} \oint {dx\over x } \ x ^{-n } \ e^{ i g\xi
 (x-1/x) \pm i g t (x+1/x)} = \( i\sqrt{ t+ \xi \over t-\xi}\)^{ \pm
 n} J_{n}(2g \sqrt{t^2-\xi^2}) \ \theta(t\pm \xi).  \la{integralforJn}
 \ee
where the contour integration goes along the unit circle, the discrete
kernel $\K$ can be expressed in terms of Bessel finctions,
   \begin{align} \la{Kmn} \mathbb{\IO}_{\ell} &= \<\ell \ | \exp \(
  {1\over 2} \sum_{m,n\ge 0} \psi_m \K_{m,n} \psi_n \) | \ell \>\!\!\>
  \ , \\
\K_{m,n} & = {g\over i} \int _{ |\xi|}^\infty d t \ X(t) \, \[\(i
\sqrt{t+\xi \over t-\xi}\)^{m-n}
 -  \(i \sqrt{t+\xi \over t-\xi}\)^{n-m}\]\no
 \\
& \qquad \times \ J_{m }(2g\sqrt{t^2-\xi^2})\,
J_{n}(2g\sqrt{t^2-\xi^2}) \, , \la{Kmndef}
\end{align}
For vacuum states of charge $\ell$, the sum in the exponential in
\re{Kmn} is effectively restricted to $m,n\ge \ell$.

\subsection{Pfaffian formula for the octagon}

 The computation of the expectation value \re{Kmn}  is 
 straightworward   and reproduces  the pfaffian formula of
 \cite{Kostov:2019auq},
  \be \la{defMM} \IO_{\ell} = { \pf \left(\begin{array}{cc} \C_{
  \gell} & 1 \\ -1 & -\K_{ \gell} \end{array}\right) = \exp\( {1\over
  2}\tr \log \left[ 1 - \C _{ \gell} \K_{ \gell} \right] \) } .  \ee
In this expression the semi-infinite matrices $ \C _{ \gell} $ and
$\K_{ \gell} $ are obtained from $\C$ and $\K$ by deleting the first
$\ell$ rows and columns.\footnote{Of course $\C_{ \gell} $ and $\C$
are identical as matrices, but considered as functions of two discrete
variables they are related by a shift by $\ell$ in both arguments.  }
For example, $\K_{ \gell} = \{ \K_{m,n}\}_{m,n\ge \ell}$.  The rhs of
\re{defMM} is defined rigorously by first truncating the semi-infinite
matrices $ \C_{ \gell}$ and $ ,\K_{ \gell} $ to $N\times N$
matrices\footnote{ If the semi-infinite matrix is truncated to a
$N\times N$-dimensional matrix, there will be an extra sign factor
$(-1)^{N(N-1)/2}$ multiplying the pfaffian.} and then taking the limit
$N\to\infty$.  The limit is convergent for any finite $g$ because
$\K_{m,n}$ decay exponentially when $m,n\to\infty$.  A more direct
derivation of the pfaffian is based on the formulation of the
expectation value as an integral over the Grassmann variables
\cite{BerezinBook},
 \be\begin{aligned} \la{pathintegral} \IO_{ \ell} &= \int \prod_{m\ge
 \ell} d \z_m d\z^*_m \ e^{ \mathcal{S}(\z,\z^*)}, \\
  \mathcal{S}(\z,\z^*)&= - {1\over 2} \sum_{m,n\ge \ell} \z _m
  \C_{m,n} \z _n + \sum_{n\ge\ell} \z^*_n\z_n +{1\over 2}
  \sum_{m,n\ge \ell} \z^*_m \K_{m,n} \z^*_n.
\end{aligned} 
 \ee

 \subsection{
 Finite pfaffian relations}

\la{section:finpfaf}

Take the operator representation of the octagon with bridge $\ell$,
eq.  \re{Kmn} and consider the right and left vacua as the result of
the action of the $\ell$ lowest fermion oscillators as in eq.
\re{defvac},
   \begin{align} \la{KImn} 
\mathbb{\IO}_{\ell} &= \< 0 | \psi_1...\psi_{ \ell-1}\ e^{{1\over 2}
\psi\K\psi}\ e^{-{1\over 2} \psi^ * \C \psi^ * } \ \psi^ *
_{\ell-1}...  \psi^ * _0 | 0\>.
   \end{align}
Hence one can obtain $\IO_\ell$ by inserting in the expectation value
for $\IO_0$ an operator creating $\ell$ pairs of fermions,
   \begin{align} \la{KImnB} 
   \mathbb{\IO}_{\ell} &= \< 0 |e^{{1\over 2} \psi\K\psi} \ B_\ell \
   |\, 0 \>\!\!\>, \quad B_\ell =\psi_1...\psi_{ \ell-1}
    \psi^ * _{\ell-1}...   \psi^ * _0 .
   \end{align}
This   can be used to derive an expression for the
  octagon  with bridge $\ell$  in terms of the expectation value of the 
  operator $B_\ell$, 
\be\begin{aligned} {\IO_\ell\over\IO_0}&= \< B_\ell\> ,
 \end{aligned}
\ee
where the expectation value of an operator $\CO$   is defined as
   \be \<\CO\> \defeq {\< 0 |e^{{1\over 2} \psi\K\psi} \ \CO \ |0
   \>\!\!\> \over \< 0 |e^{{1\over 2} \psi\K\psi} \ |0 \>\!\!\> } .
   \ee

As any expectation value of free fermions, $\<B_\ell\>$ is equal to
the pfaffian of the two-point correlation functions of the fermions
involved.  A direct calculation gives, for $j,k=0,1,2,...$,
 \be 
 \begin{aligned}
 \<  \psi^ * _j\psi ^*_k\> & = - [\K(1+ \RR)]_{j,k} , \\
 \< \psi_j\psi ^*_k\> & = \quad [ 1+ \RR ]_{j,k} ,\\
 \< \psi^ * _j\psi _k\> & = \ \ - [1+\RR]_{ k,j} , \\
 \< \psi_j\psi _k\> &
 =
\quad  [(1+\RR)\C]_{j,k}\,
\la{DrProp} 
\end{aligned}
  \ee
where
\be \la{defRR} \RR= {\C\K\over 1-\C\K} .  \ee
The matrix of all correlators is the inverse of the quadratic form in
the representation as integral over grassman variables, as it should,
\be
\la{inverserelprops} \left(\begin{array}{cc} (1+\RR)\C & 1+\RR \\
- (1+\RR)^{\mathrm{T}} & - \K (1+\RR) \end{array}\right) =
\left(\begin{array}{cc} - \K & -\I \\ \I &\ \C
\end{array}\right)^{-1} . 
\ee
Now we can express the ratio $\IO_\ell/\IO_0$ as an $2\ell\times
2\ell$ pfaffian
 \be \la{finitepff} {\IO_\ell\over \IO_0} =( -1)^{{\ell(\ell-1)\over
 2}}\ \pf \[ \left(\begin{array}{cc} (1+\RR)\C & 1+\RR \\ -
 (1+\RR)^{\mathrm{T}} & - \K (1+\RR) \end{array}\right)_{\!\! <\ell\ }\]. 
  \ee
 Here we introduced the symbol $\XX\lell$ which represents the
 truncation of the semi-infinite matrix $\XX$ to an $\ell\times\ell$
 matrix $\{\XX_{m,n}\}_{0\le m,n <\ell}$.
 The truncation is applied to all four blocks of the matrix.
 
We have checked the finite pfaffian relation \re{finitepff} 
for $\ell\le 4$ up to
$g^{16}$.  As anotherconsistency check let us consider the limit
$\ell\to\infty$ where $\IO_\ell\to 1$.  Then after taking into account
\re{inverserelprops}, the identity \re{finitepff} reproduces the
original pfaffian formula \re{defMM} for $\ell=0$.

  An obvious generalisation of \re{finitepff} relates two octagons
  with bridges $\ell <\ell_1$,
 \be \la{defMMl1l} {\IO_{\ell_1}\over \IO_{\ell }} = (
 -1)^{{(\ell_1-\ell)(\ell_1-\ell-1)\over 2}}\ \pf
 \[\left(\begin{array}{cc} (1+\RR_\gell )\C_\gell & \quad 1+\RR_\gell \\
 & \\ - 1-\RR^\mathrm{T} _{ \gell }&
  \quad  - \K\gell(1+\RR\gell) \end{array}\right)_{\!\!\!\!  <\ell_1}\ \]
.
 \ee
In particular, the octagons with bridges $\ell$ and $\ell+1$ are
related as
\be
\la{diagonalformula} {\CO_{\ell+1}\over\CO_\ell}=
 \[ {1\over 1- \C_{ \gell}   \K_{ \gell} }\]_{\ell,\ell}  
\ee
which provides a factorised  form for  the  relation 
\re{finitepff}.

 \section{The similarity transformation  }
 \la{section:similaritytr}

 \subsection{The original and the simplified octagon kernels}
 
In this section we give explicit expression for the similarity
transformation relating the original and the simplified octagon
kernels, which corresponds to the canonical transformation
\re{changeofbas} of the fermion oscillators.  The operator
representation based on the new set of oscillators has the advantage
that it preserves the $ U(1)$ charge and therefore leads to a
determinant instead of a pfaffian.
 
  In this subsection we remind the definition of the two kernels.
It is convenient to change the variables
 in \re{Kmndef} as
 \be \la{defst} \xi \equiv {\s\over g},\quad t\equiv {1\over g} \sqrt{
 \t^2+ {\s^2}}.  \ee
 so that the integration now spreads on the whole positive real axis
 and the dependence on the 't Hooft coupling is carried only by the
 weight function $\X$.  In the new variables, the weight function
 takes the form
   \be \la{defX} \chi(\t,\s ) \equiv X({\sqrt{\t^2+\s^2} \over g}) = {
   \cos \phi - \cosh\vp \over \cos\phi - \cosh {\sqrt{\t^2+\s^2} \over
   g}} \, \ee
and the integral formula for the matrix elements \re{Kmndef} becomes
\be
  \begin{aligned}
 \la{originalkernelbis} \K_{m,n} \ &= 2 \int _{ 0}^\infty d \t \
 \chi(\t,\s ) \ \CK_{m,n}( \t; \s) \qquad (m,n\ge 0),
   \end{aligned}
      \ee
 with
  \begin{align} 
   \CK_{m,n}(\s, \t) &= \Pi_{m-n}(\s/\t)\ J_{m }(2\t)\ J_{n}(2\t),
   \la{defCKmn} \\
 \no
 \\
 \Pi_{n }(z) \ &\defeq\ { \frac{ i^ {n } \left(\sqrt{ z^2
 +1}+z\right)^{n }-i^{-n } \left(\sqrt{z^2+1}-z\right)^{n } }{2i
 \sqrt{z^2+1}} }
 =
  - \Pi_{-n }(z)
  .  \la{defPin1}
\end{align} 
Importantly, $\Pi_n$ is a polynomial, \be \Pi_0(z)=0, \ \Pi_1(z)= 1,
\Pi_2(z) = 2 i z, \ \Pi_3(z)= - 1 - 4 z^2, \ \ \text{etc}.  \ee
It equals the $(n-1)$-th Chebyshev polynomial of second kind with
imaginary argument.  We give the explicit expression for the
coefficients of this polynomial, which will be needed in the
following,
\be
 \la{expCheby}
\begin{aligned}
 \Pi_n(z) &= U_{n-1}(iz) = \sum _{p =0} ^{n-1} \sin{ \textstyle \pi {
 n -p\over 2}}\ A_{n}^{(p)}\ {(-i z)^p\over p!} \qquad \qquad (n\ge 1)
 \end{aligned} 
  \ee
with
 \be
\la{defAmn}
\begin{aligned}
A_{n } ^{(p)} = (-2)^p \frac{ \Gamma \left[\frac{1}{2} ( n +1+p
)\right]}{ \Gamma \left[\frac{1}{2} ( n +1-p )\right]}.  \end{aligned}
\ee
Summarising, the integrand in \re{originalkernelbis} is given by
a sum of products of Bessel functions,
 \begin{align} 
\la{transP} \CK_{m,n}( \t; \s) = \sum _{j =0} ^{|m-n|-1} {1-(-1)^ {m-
n-j}\over 2} \ A_{m-n}^{(j)}\ {(i \s)^j\over j!} \ {i^{m-n-1+j}\,
J_{m}(2\t)\ J_{n}(2\t)\over \t^j} \,.
\end{align}

The octagon kernel \re{originalkernelbis} depends on the cross ratios
of the spacetime coordinates (the parameters $\s $ and $\phi $)
through the weight function $\chi$ and also through the polynomials
$\Pi_{m-n}(\s/\t)$.  It was noticed \cite{Kostov:2019auq} that the
second dependence is redundant in the sense that only the constant
terms $\Pi_{m-n}(0) = \sin({m-n\over 2}\pi) $ of these polynomials
contribute.  Based on this observation, it was conjectured that in the
pfaffian formula \re{defMM}, the kernel $\K$ can be replaced with a
simplified kernel whose matrix elements $\KI_{m,n}$ are real and
vanish if $m$ and $n$ have the same parity.  The last property implies
that the pfaffian \re{defMM} can be written as a determinant.  The
simplified kernel was found as a perturbative series in
\cite{Kostov:2019auq} and in integral form in \cite{Belitsky:2020qrm},
  \be  \la{improvedkernel}
  \begin{aligned} 
     \KI _{m,n} &= 2 \int _{ 0}^\infty d \t \ \chi(\t ,\s) \ \
     \CKI_{m,n}(\t) \qquad (m,n\ge 0) \\
      \CKI_{m,n}(\t)&= \sin\( \textstyle{ m-n\over 2}\pi\) \
      J_{m}(2\t)\ J_{n}(2\t) ,
 \end{aligned} 
 \ee 
 where $\chi(\t,\s) $ is the  weight
 function  defined in \re{defX}.  
 
The conjecture of \cite{Kostov:2019auq} states, with the
interpretation of the bridge we adopted here, that for any $\ell\ge
0$, the matrices $\C_{ \gell} \K_{ \gell} $ and $\C_{ \gell} \KI_{
\gell} $ are related by a similarity transformation.  (We remind that
$\XX_{\gell}$ denotes the matrix $\XX$ with its first $\ell$ rows and
columns deleted.)  This is equivalent to claiming that there exists a
symplectic transformation preserving $\C$ and relating $\K_{ \gell} $
and $ \KI _{ \gell} $,
 \be \la{similaritytrgen} \C_{ \gell} \K_{ \gell} = \UU_\ell ^{-1}\C_{
 \gell} \KI_{ \gell} \UU_\ell \quad \Leftrightarrow\quad
 \begin{cases}
 \K_{ \gell} = \UU_\ell ^{\mathrm{T}} \KI_{ \gell} \UU_\ell , & \\
    \C_{ \gell} = \UU_\ell \C_{ \gell} \UU_\ell^{\mathrm{T}} .&
\end{cases}
   \ee
In terms of the ensemble of fermions, the above statements mean,
first, that the operators in the expectation values \re{operoct} and
\re{operoctb} are related by the canonical transformation
\re{changeofbas}, and second, that the canonical transformation in
question leaves the bra and ket vacua of charge $\ell$ invariant.

 \subsection{ Explicit solution for the similarity transformation as a
 power series }

 The solution for the matrix $\UU_\ell$ in \re{similaritytrgen} is not
 unique.  We found a particular solution of the first equation
 \re{similaritytrgen} in the form of a power series in $\s$,
\be
\la{Nrel1}
\begin{aligned}
 \UU_\ell & = \sum_{ p = 0} ^{\infty } {(-i \s)^p \over p !} \ \(\C_{
 \gell} \MM _{ \gell} \)^p \QQ^{(p)} _{\ell}
  \end{aligned}
\ee
 where the diagonal matrices $\MM_{ \gell} $ and $ \QQ_\ell^{(p)}$ are
 defined as\footnote{ The lowest matrix element $[\MM_{ \gell}
 ]_{\ell,\ell}$ is singular for $\ell=0$, but it does not appear
 neither in \re{Nrel1} nor in the matrix relations further on. 
 }
\begin{align}  
  \la{defM} \MM_{ \gell} &= \mathrm{diag}\left \{{1\over
  n}\right\}_{n\ge \ell}= \mathrm{diag} \left\{{1\over \ell}, {1\over
  \ell+1}, {1\over \ell+2},...  \right\} \\
   \la{defcoefY} \QQ_\ell^{(p )} &= \mathrm{diag} \left\{
   \th_{n+1-p-\ell}\ \( \a_{n-\ell-p} \ A_{n-\ell }^{(p)} +
   \a_{n-1-\ell-p} (-1)^p \ B_{ n-\ell } ^{(p)} \) \right\}_{n\ge
   \ell}.
    \end{align}
  Here
  \be \la{defalpha} \a_k \equiv \hf(1- (-1)^k), \qquad \th_k =
     \begin{cases}
	  0& \text{if} \ \ k\le 0, \\
   1   & \text{ if} \ \ k >   0 
\end{cases}
       \ee
 the coefficients $ A_{n} ^{(p)} $ 
are    defined above in \re{defAmn},
and the coefficients $ B_{m}^{(p)
 } $ are given by 
	\be 
	\la{relB-A} B_{m}^{(p)} =m (-1)^{p-1} A_m^{( p-1)} = \frac{2^{p-1} m
	\Gamma \left(\frac{1}{2} (m +p)\right)}{\Gamma \left(\frac{1}{2}
	(m +2-p)\right)} \,.  \ee
 These coefficients appear in  
  the Taylor expansions
 \be
 \la{ABcoef}
\begin{aligned}
\frac{\left(\sqrt{z^2+1}-z\right)^m}{\sqrt{z^2+1}}&=\sum_{p\ge 0} A^{(
p )}_{m } {z^p \over p!} , \quad \quad \left(\sqrt{z^2+1}+z\right)^m=
\sum_{p\ge 0} B_{m}^{(p) } {z^p \over p!} \,.
 \end{aligned}
\ee

For fixed $m ,n\ge \ell$, the matrix element $(\UU _\ell)_{m,n}$ is a
polynomial in $\s$ of degree $n-\ell$ for $m-\ell$ even, or of degree
$n-1-\ell$ for $m-\ell $ odd.  The coefficients of this polynomial depend
explicitly on the bridge length $\ell$.  The lowest matrix elements $
\ell\le m,n\le \ell+ 3$ are
\be 
\UU _\ell=\left(
\begin{array}{cccccc}
 1 \ \ & \frac{i \sigma }{\ell +1} \ \ & -\frac{2 \sigma ^2}{(\ell
 +1)(\ell+2)} \ \ & -\frac{4 i \sigma ^3}{(\ell +1) (\ell +2) (\ell
 +3)}
 \ \ &* 
  \\
 0 \ \ & 1 \ \ & \frac{2 i \sigma }{\ell +2} \ \ & -\frac{4 \sigma
 ^2}{ (\ell+2)(\ell+3)} \ \ & * \\
 0 \ \ & -\frac{i \sigma }{\ell +1} \ \ &1+ \frac{4 \sigma ^2}{
 (\ell+1)(\ell+3)} \ \ & \frac{3 i \sigma ( \ell+1) (\ell +4 ) + 12 i
 \sigma ^2 }{(\ell +1) (\ell +3) (\ell +4)} \ \ & * \\
 0 \ \ & 0 \ \ & -\frac{2 i \sigma }{\ell +2} \ \ &1+ \frac{8 \sigma
 ^2}{ (\ell+2)(\ell+4)} \ \ & * \\
 0 \ \ & 0 \ \ & -\frac{2 \sigma ^2}{ (\ell+2)(\ell+3)} \ \ & -\frac{3
 i \sigma \left(4 \sigma ^2+\ell (\ell +7)+10\right)}{(\ell +2) (\ell
 +3) (\ell +5)}  \ \ & * \\ 
 0\ \ &0 \ \ &0\ \ &* \ \ &*
\end{array}
\right)
.
\ee
 We give the idea of the derivation of the symplectic transformation
 in appendix \ref{Appendix:C}.  The proof is based on a linear
 relation between  
 $\K_{ \gell} $ and $\KI_{ \gell} $\,,
\be\la{Nrelmn}
\begin{aligned}
 \K _{m,n} & =\sum _{k= 0}^{m-n-1} {( i \s)^k\over k!} \a_{m-n-k}\ \
 A^{( k )}_{m-n }[(\MM \C)^k \K^\circ]_{m ,n}\qquad \ \ \ \ \ (m > n ) \\
 \K _{m,n} & =- \sum _{k=0}^{n-m-1} {(- i \s)^k\over k!} \a_{m-n-k}\
 A^{( k )}_{n-m }[(\K^\circ (\C\MM)^k]_{m ,n}\qquad (m <n),
  \end{aligned}
 \ee
which follows from the expansion \re{transP} and the recurrence
relation for the Bessel functions
 \be
 \la{recur0}
 \begin{aligned}
 J_{m+1} (2\t) + J_{m-1} (2\t) = m\ { J_m(2\t)\over \t} ,
 \end{aligned}
\ee
see appendix \ref{Appendix:A}. 
Concerning the second relation in \re{similaritytrgen}, we checked
that it is satisfied by the series \re{Nrel1} for the first several
orders in $\s$, but we do not know how to prove it analytically in
general.  In the next subsection we give another form of the solution
\re{Nrel1} for $\ell=0$, for which this property comes out naturally.

 As we mentioned before, the similarity transformation is not unique,
 and another solution was independently obtained by Belitsky and
 Korchemsky \cite{ Belitsky:2020qir}.  In appendix \ref{Appendix:B} we
 re-derive their result as a solution of an ordinary differential
 equation describing the operator flow connecting $\K$ and $\KI$.

 \subsection{ Exponential form of the solution for $\ell=0$}
 \la{ssec:exp}

 When $\ell=0$, the solution \re{Nrel1} for the similarity
 transformation can be written in a quasi exponential form,
\be
\la{Fullexponentialsolution}
  \UU _{\ell=0}
=\sum_{j=0}^\infty
 \[\(\PP _\even\ 
e^{-{1\over 2} \s^2 \C\MM\SS} + \PP _\odd \ e^{-{1\over 2} \s^2
\SS\C\MM} \)e^{i\s\C}\] _j \PP ^{(j)}.  \ee
Here $[...]_j$ denotes the coefficient of the power $\s^j$ in the
expansion of the expression in the brackets, $\MM\equiv \MM_{\ell=0}$
is given by \re{defM}, the matrix $\SS$ is defined as
\be\begin{aligned} \la{defS} \SS_{m,n} = \d_{m+1,n}+\d_{m ,n+1} \qquad
(m,n\ge 0)\ ,
\end{aligned}
\ee
$ \PP^{(j)} $ is as in \re{defcoefY} the projector to the matrices
with the first $j$ columns vanishing,
  \be
\la{defProj}
 \begin{aligned}
 \PP^{(j)} & = \mathrm{diag} \{ \theta_{n+1-j } \}_{n\ge 0}\, ,
        \end{aligned}
\ee
and $\PP_{\even }$ and $\PP_\odd $ are the projectors respectively to
the even and odd subsets,
\be \PP _\even =\mathrm{diag}\{ \a_{m+1}\}_{ m\ge 0}\, , \ \ \PP _\odd
=\mathrm{diag}\{ \a_m\}_{ m\ge 0}\, \ee
with $\a_k$ given by \re{defalpha}.

To get some intuition on the origin of the two exponential factors in
\re{Fullexponentialsolution}, let us write the simplified kernel for
$\ell=0$ in $x$-representation,
\be \begin{aligned} \la{intrepKKI} \overset{_\circ} K  (x,y) &
 = 
\sum_{m,n \in \IZ} x^{m} y ^{n} \, \KI_{m,n} = 2 \int_0 ^\infty d\t\
\chi(\t,\s)\ \sin \[{\textstyle \(x+\frac{1}{x}- y-\frac{1}{y}\right) }\t
\],
\end{aligned}
\ee
and compare it with the original kernel \re{defKxy}, written in
terms of the variables \re{defst},
\be
 \begin{aligned}
\la{Kxyinnewvars} K (x,y) &= 2 \ e^{i \s \, (x -{1\over x} +y- {1\over
y})} \int_0 ^\infty d\t\ \chi(\t,\s)\ {\t\over \sqrt{\t^2+\s^2}} \sin \[
{\textstyle \(x+\frac{1}{x}- y-\frac{1}{y}\right)}\sqrt{\t^2+\s^2}\].
\ \ \ \
\end{aligned}  
\ee
 The first expression is obtained from the second by setting $\s=0$ {\ everywhere but  in the factor $\chi(\t,\s)$.
 In \re{Fullexponentialsolution}, the right exponential factor
 $e^{i\s\C}$ accounts for the factor $ e^{i \s \, (x -{1/ x} +y- {1/
 y})} $ in \re{Kxyinnewvars}.  Indeed, in $x$-representation, the
 operator $\C$ acts as a multiplication by $x-1/x$.  The second factor
 in \re{Kxyinnewvars} originates from the $\s^2$-dependence of the
 integrand of \re{Kxyinnewvars}.  The latter expands as a series in
 $\s^2$, with the constant term given by the integrand of
 \re{intrepKKI}.

 \section{ Determinant identities    }
\la{section:dets}

Since in the simplified kernel \re{improvedkernel} the matrix elements
with the same parity vanish, the $2\ell\times 2\ell$ pfaffians in the
finite pfaffian formulas obtained in section \ref{section:finpfaf} can
now be written as $\ell\times \ell$ determinants.  It turns out that
these determinants can be simplified further and written as
determinants of approximately twice less size.  More precisely, for
$\ell= 2m-1$ and $\ell=2m$, the ratio $\IO_\ell/\IO_0$ is an $m\times
m$ determinant.

 To obtain the reduced determinant identities, we first notice that in
 the Fock space representation \re{operoctb} the exponents are
 bilinear forms of the even and odd modes,
  \be
    \begin{aligned}
  \la{Kmnevenodd} \mathbb{\IO}_{\ell} &= \<\ell \ | \exp \(
  \sum_{j,k\ge 0} \psi_{2j+1} \KI_{2j+1,2k} \, \psi_{2k} \) \ \exp \(-
  \sum_{j,k\ge 0} \psi^ * _{2j} \C_{2j ,2k+1 } \, \psi_{2k+1}^* \)|
  \ell \> .
  \end{aligned} 
  \ee
 We will show that, depending on the parity of $\ell$, one can
 eliminate either the even or the odd modes from the expectation value
  taking into account  the identification
 \re{propagatorF}.

  \subsection{ Even   bridge }
  \la{ss:octagonasdeterminant}

Let us assume that the length of the bridge is even, $\ell=2m$.  In
the operator expression \re{Kmnevenodd}, we can commute all the even
modes of $ \psi^ * $ to the left and all the even modes of $\psi$ to
the right until they both are annihilated by the corresponding vacua.
As a result we obtain an operator expression only in terms of the odd
modes,
  \be
    \begin{aligned}
  \la{Kmnevenodd2} \mathbb{\IO}_{\ell= 2m} &= \sum
  _{n=0}^\infty{(-1)^n\over n!} \sum_{^{j_1,...,j_n \ge
  m}_{k_1,...,n_n \ge m}} \ \<\ell \ | \prod_{a=1}^n \psi_{2j_a+1}
  \prod_{a,b=1}^n \IK ^{\odd\odd}_{j_a,k_b}
  \prod_{b=1}^n\psi^*_{2k_b+1} | \ell \> \\
   &=
   \< \ell| \, \antinor \exp\(-\sum_{k,j\ge 0} \psi_{2j+1} \,
    \IK^{\odd\odd}_{j,k} \, \psi^*_{2k+1}\)  \antinor \, |\ell
   \>.
  \end{aligned} 
    \ee
In the last line $\antinor\ \antinor$ denotes the anti-normal ordering
where all $\psi^*$ are on the right of all $\psi$.  By $\IK^{\odd\odd}
$ we denoted odd-odd diagonal block of the matrix $ \KI\C$,
  \be\la{defIKeeooter} \IK ^{\odd\odd}_{j,k} \defeq [\KI \C] _{2j+1 ,
  2k+1} \, , \ee
whose matrix elements are given explicitly by
\be
\la{oddkerneldiscret}
\begin{aligned}
   \IK ^{\odd\odd} _{i,j} & = 2(2j+1 )\, (-1)^{i-j } \int _{ 0}^\infty
   d \t \ \chi(\t, \s)\ {J_{2i+1 } (2\t) \ J_{2j+1} (2\t) \over \t} \qquad
   (i,j\ge 0) .
 \end{aligned}
 \ee
  To obtain the rhs of \re{Kmnevenodd2} we used the identity
  \be
    \la{defIKeeoobis} 
    (\KI_\gell \C_\gell ) _{2j+1
  ,
  2k+1}
=  ([\KI\C]_\gell)_{2j+1,2k+1}
  \qquad (\ell= 2m)  \ee
 which follows from the fact that the matrix $\C$ is quasi-diagonal.
 Evaluating the expectation value with the correlators \re{ccrf}, we
 obtain the determinant formula for the octagon
\be \la{improveddetb} \IO_{\ell= 2m}  = \det [\( 1-
\IK^{\odd\odd}\)_{\ge m}]. \ee

Now we would like to evaluate the ratio of the octagons with $\ell=2m$
and $\ell=0$ as an expectation value, as in section
\ref{section:finpfaf}.  The identity \re{defIKeeoobis} also guarantees
that the even fermion modes in the vacuum states can be removed
without altering the result,
   \be\begin{aligned}
   \la{oddvac} | \ell \> &\ \to \ \psi^{* }_{2m-1} \psi^{* \
   }_{2m-3}\cdots \psi^{* }_1 |0 \> \equiv |m,\odd\>\, , \\
   \< \ell |&\ \to\  \<
  0  | \psi _1 \psi _3\cdots\psi _{2m-1}   
 \equiv \<m, \odd|
   \end{aligned}
     \qquad \qquad (\ell= 2m, \ m\ge 1).\ee

We express the octagon with bridge $\ell= 2m$ as the result of the
insertion of $m$ pairs of odd fermionic modes, and divide by the
octagon with bridge zero,
 \be
 \begin{aligned}
 {\IO_{2m}\over \IO_0}&= {\<0  | \psi _1 \psi _3\cdots\psi _{2m-1} \
 \antinor e^{\psi \IK^{\odd\odd} \psi^*} \!  \antinor\ \psi^{*
 }_{2m-1} \psi^{* \ }_{2m-3}\cdots \psi^{* }_1 |0  \> \over \<0  |
 \antinor e^{\psi \IK^{\odd\odd} \psi^*} \antinor |0  \> } \\
 &  \equiv \< \prod_{j=0}^{m-1}
 \psi _{2j+1} \prod_{j=0}^{m-1}\psi^{* }_{2j+1} \>.  
  \end{aligned}
  \ee
The expectation value is equal to the determinant of the two-point
correlators
 \be \< \psi _{2j+1} \psi^{* }_{2k+1} \> = \d_{j,k}+ \IR
 ^{\odd\odd}_{j,k}, \ee
where the semi-infinite matrix $\IR^{\odd\odd}$ is related to $\IK
^{\odd\odd}$ by
    \be \la{defRoo} (1+ \IR^{\odd\odd})(1- \IK^{\odd\odd})=1.  \ee
 Hence the ratio $\IO_{2m}$ and $\IO_0$ is an $m\times m$ determinant,
\be \la{Reedeteven} {\IO_{2m}\over \IO_0} =\det\[ (1+ \IR
^{\odd\odd})_{< m}\]
.
\ee
Since $ \IO_\ell\to 1$ when $\ell\to\infty$, eq.  \re{Reedeteven}
reproduces in the large $\ell$ limit the determinant formula for the
octagon with zero bridge,
 \be \la{detformoct}
 {1\over \IO_0}= \det\[ 1+\IR  ^{\odd\odd}\]
 = {1\over \det[1-\IK^{\odd\odd}]}.  \ee

Since we never used the specific form of $\IK^{\odd\odd}$, the 
identity \re{Reedeteven}  is in fact an identity in the linear algebra.\footnote{We
thank G. Korchemsky for making this point.} Namely, for any
non-singular matrix $A$,
\be \la{linearalgebra} {\det \[A_{\ge k}\]\over \det A} =     \det\[\{ (A^{-1})_{i,j}
\} _{i,j=0,...,k-1} \],  \qquad A=\{ A_{i,j}\}_{i,j=0,...,k-1} 
\, .  \ee
Indeed, the octagon with bridge $2m$ is given by the determinant of
the kernel $\IK^{\odd\odd}$ with the first $m$ rows and columns
deleted,
\be \la{improveddetbodd} \IO_{\ell= 2m} = \< m,\odd | \antinor e^{-
\psi \IK^{\odd\odd} \psi^{* } } \antinor |m,\odd\>\, = \det [\( 1-
\IK^{\odd\odd}\)_{\ge m}], \ee
and \re{Reedet} follows from the general identity \re{linearalgebra}.

  \subsection{Odd bridge}
  
In a similar way, in the case $\ell= 2m-1$ one can eliminate all odd
modes in \re{Kmnevenodd}.  As a result we obtain an operator
expression only in terms of the even modes,
 \be \begin{aligned}
	\la{Kmnevenodd22} \mathbb{\IO}_{\ell= 2m-1} &= \sum
	_{n=0}^\infty{(-1)^n\over n!} \sum_{^{j_1,...,j_n \ge
	m}_{k_1,...,n_n \ge m}} \ \<\ell \ | \prod_{a=1}^n \psi_{2j_a }
	\prod_{a,b=1}^n \IK ^{\even\even}_{j_a,k_b}
	\prod_{b=1}^n\psi^*_{2k_b } | \ell \> \\
   &= \< \ell| \, \antinor \exp\(-\sum_{k,j\ge 0} \psi_{2j} \,
   \IK^{\even\even}_{j,k} \, \psi^*_{2k}\) \antinor \, |\ell \>.
  \end{aligned} 
    \ee
   The matrix elements of 
  \be
  \IK ^{\even\even}_{i,j} \equiv [\KI\C]_{2j , 2k }
  \ee
  are given explicitly by ($j,k\ge 0$)
\be
\la{kernelevenbr}
\begin{aligned}
  \IK ^{\even\even} _{i,j} & = 2 \int _{ 0}^\infty d \t \ \chi(\t, \s)\
  \[(1-\d_{j,0}) (-1)^{i-j } 2j {J_{2i } (2\t)
  J_{2j } (2\t) \over \t} +\d_{j,0} (-1)^j J_{ 2i}J_{  1}\].
 \end{aligned}
 \ee
Thanks to the identity $[\KI_\gell \C_\gell ]_{2j,2k} =
[[\KI\C]_\gell]_{2j,2k} $ for $\ell$ odd, we can eliminate the odd
modes also from the left and the right vacuum and formulate the
octagon as an expectation value in the ensemble of the even
oscillators,
  \be
  \begin{aligned}
  \la{improvedOpd} \IO_{\ell= 2m-1} &= \< m,\even | \antinor e^{ -
  \psi^\even \, \IK^{\even \even } \psi^{* \even } } \antinor |m,\even
  \>\, = \det\[ \(1 - \IK^{\even\even}\)_{\ge m} \] \qquad (m\ge 1).
  \end{aligned}
  \ee
The bra and ket vacuum states here are the $m$-charged vacua for the
even modes,
  \be\begin{aligned}
  \la{evenvac}
   |m, \even \> = \psi^{* }_{2m-2} \psi^{* }_{2m-4
  }\cdots \psi^{* }_{0 } |0\>\, , \ \ \< m,\even |= \< 0| \psi _{0}
  \psi _{2} \cdots\psi _{2m-2} .  \end{aligned}\ee

Again, the ratio $\IO_{2m-1}/\IO_0$ can be computed as an expectation
value of $m$ fermion pairs which gives an $m\times m$ determinant
 \be \la{detreloddbr} {\IO_{2m-1}\over \IO_0} = \< \prod_{j=0}^{m-1}
 \psi^\even_{2j} \psi^{* \even}_{2j }\> = \det\[\( 1+\IR ^{\even\even}
 \)_{< m}\], \ee
where the two-point correlator $\< \psi^\even_j \psi^{* \even}_k\>
=\d_{j,k}+ \IR ^{\even\even}_{j,k} $ is related to the even-even
kernel \re{kernelevenbr} by\footnote{The relation with the full
resolvent is $\IR^{\even,\even}_{j,k} =[\C\KI/(1-\C\KI] _{2j,2k}$ and
$\IR^{\odd\odd}_{j,k} =[\C\KI/(1-\C\KI] _{2j+1,2k+1}$.}

   \be (1+ \IR ^{\even\even})(1- \IK^{\even\even})=1.  \ee

Taking the large $\ell$ limit of \re{detreloddbr}, we reproduce the
determinant formula for the octagon with zero bridge in terms of the
even kernel.  Thus the octagon with zero bridge can be expressed as a
determinant in either of the sectors
 %
  \be \la{oct0eo} \IO_0= \det\( 1- \IK^{\odd\odd}\)=\det\( 1-
  \IK^{\even\even}\).  \ee

We have checked that the finite determinant identities obtained in
this section are fulfilled within the weak coupling expansion of the
octagon.  Let us stress that although the expressions
\re{improveddetb} and \re{improvedOpd} look differently, they are both
identical to the determinant of the odd block which was the starting
point for the studies in
\cite{Belitsky:2019fan,Belitsky:2020qrm,Belitsky:2020qir}.  The
difference is in the interpretation: in
\cite{Kostov:2019auq,Belitsky:2019fan,Belitsky:2020qrm,Belitsky:2020qir}
the bridge appeared as a parameter while here it is the effect of
truncating the semi-infinite matrix.

 \subsection{ The octagon as a Fredholm determinant of a holomorphic
 kernel }
  \la{sec:continuous}

We will show that the operator representation with the simplified
kernel \re{operoctb} can be expressed as the expectation value of an
exponential operator which commutes with the $U(1)$ charge.  For that
we will represent the ordered exponentials in the expectation values
\re{Kmnevenodd2} and \re{Kmnevenodd22} as ordinary exponentials.  This
can be done at the expense of extending the sum in the exponents to
all possible modes, positive and negative, after having extended the
semi-infinite matrix $\IK^{\a\a}$ to a doubly infinite matrix.

Let us choose an even bridge $\ell = 2m$ so that $\a=\odd$.  The
expectation value \re{Kmnevenodd2} can be expressed as that of an
ordinary exponential as
 \be\begin{aligned} \la{octxrepdiscrodd} \IO_{ \ell= 2m} &=\<\ell |
 \exp\(-\sum_{j,k\ge 0}\IK_{j,k} ^{\a\a} \ \( \psi_{2j+1} +
 \psi_{-2j-1} \) \(\psi^*_{2k+1} - \psi^*_{-2k-1}\)\)| \ell\> \\
&= \< \ell |  \exp\(-\sum_{j,k\in \IZ}\IK_{j,k} ^{\a\a}
\  \psi_{2j+1}\ \psi^*_{2k+1}\)| \ell\>,
 \end{aligned} 
  \ee
where in the second line the octagon kernel is extended to negative
values of the indices by the symmetries $\IK^{\odd\odd}_{j,k} =
\IK^{\odd\odd}_{-j-1,k} =-\IK^{\odd\odd}_{j, -k-1}$.  To illustrate
why the negative modes are necessary, compare the quadratic terms in
the expansion of \re{octxrepdiscrodd} with that of \re{Kmnevenodd2},
\be
 \begin{split}
 \re{octxrepdiscrodd} \ \Rightarrow &=\sum_{j,k\ge 0} \sum_{i,r\ge 0}
 ( \IK^{\odd\odd}_{i, j}\IK^{\odd\odd}_{k,r}\, \<0| \psi_{2i+1}
 \psi^*_{2j+1} \psi_{2k+1}\psi^*_{2r+1}|0\> \ \\
& +\IK^{\odd\odd}_{i, -j-1}\IK^{\odd\odd}_{-k-1,r} \, \<0| \psi_{2i+1}
\psi^*_{-2j-1} \psi_{-2k-1}\psi^*_{2r+1}|0\> )\\
&= \sum_{ i,j\ge 0} \IK^{\odd\odd}_{i, i}\IK^{\odd\odd}_{j, j} +
\sum_{ i,j\ge 0} \IK^{\odd\odd}_{i, -j-1}\IK^{\odd\odd}_{-j-1, i} \\
&= \sum_{ i,j\ge 0} \IK^{\odd\odd}_{i, i}\IK^{\odd\odd}_{j, j} -
\sum_{ i,j\ge 0} \IK^{\odd\odd}_{i, j}\IK^{\odd\odd}_{j, i} \
\Leftarrow \ \re{Kmnevenodd2}.
 \end{split}
\ee

By rewriting the exponent in \re{octxrepdiscrodd} as a double contour
integral, we get
\be\begin{aligned} \la{octxrepa} \IO_{ \ell = 2m} &= \<\ell |
\exp\(- {1\over 4\pi^2} \oint_{\mathcal{C}} {dx\over   x}\oint _{\mathcal{C}^*}
{dy\over y} \psi(x)\slK^{\odd\odd}(x,y) \psi^{*} (y) \) |
\ell\>, \end{aligned} \ee
 where the contour $\mathcal{C}^*$ contains the origin and is
 contained in the contour $\mathcal{C}$, and the holomorphic kernel is
 given by
  \be\begin{aligned}
  \la{LAurentexp}
 \slK^{\odd\odd}(x,y) &= \sum_{i,j\in\IZ}\IK ^{\odd\odd} _{i,j}\ x^{2i+1}y^{-2j-1}
 \\
&= -2 \left( y-\textstyle{\frac{1}{y}}\right) \int_0^ \infty d\t
\chi(\t,\s)\sin \left[\tau \left(x+
\textstyle{\frac{1}{x}}\right)\right] \ \cos \left[\tau
\left(y+\textstyle{\frac{1}{y}}\right)\right] \\
& =-     2   \int_0^  \infty  {d\t\over \t}\chi(\t,\s)
\ \sin \left[\tau  \left(x+
\textstyle{\frac{1}{x}}\right)\right]
\  y\p_y  \sin  \left[\tau  \left(y+\textstyle{\frac{1}{y}}\right)\right] .
\end{aligned} 
\ee

In a similar way, for $\ell $ odd we can write
 \begin{align} \la{octxrepabis}
\IO_{ \ell= 2m-1} &= \<\ell| \exp\(-\oint_{\mathcal{C}} {dx\over 2\pi
i x}\oint _{\mathcal{C}^*} {dy\over 2\pi i y} \psi(x)
\slK^{\even\even}(x,y) \psi^{*} (y) \) | \ell\>, \\
 \no
 \\
\no \slK^{\even\even}(x,y) &= \sum_{i,j\in\IZ}\IK ^{\even\even}_{i,j}\
x^{2i }y^{-2j } \\
&= 2 \left(y-\textstyle{\frac{1}{y}}\right) \int_0^ \infty d\t
\chi(\t,\s)\cos \left[\tau \left(x+
\textstyle{\frac{1}{x}}\right)\right] \ \sin \left[\tau
\left(y+\textstyle{\frac{1}{y}}\right)\right] \no \\
&= 2   \int_0^ \infty{d\t\over \t}
\chi(\t,\s)\cos \left[\tau \left(x+ \textstyle{\frac{1}{x}}\right)\right]
y\p_y  \cos \left[\tau \left(y+\textstyle{\frac{1}{y}}\right)\right].
\la{Kernelxeven}
\end{align}  
The sum of two kernels, \re{LAurentexp} and \re{Kernelxeven}, gives
the $x$-representation of the full operator $ \IK \equiv \KI\C=
\IK^{\even\even}\otimes \IK^{\odd\odd}$.  The factor $y-1/y$ results
from the action of the operator $\C$ which diagonalises in the
$x$-space, and the rest reproduces the rhs of \re{intrepKKI}.

The operator representaton of the octagon in $x$-space, eqs.
\re{octxrepa}-\re{LAurentexp}, was derived for even bridge $\ell= 2m$.
It is possible to extend each of the representations,
\re{octxrepa}-\re{LAurentexp} and \re{octxrepabis}-\re{Kernelxeven},
to any value of the bridge length, odd and even.\footnote{The
derivation of \re{octxrepa} for odd bridge and of \re{octxrepabis} for
even bridge is slightly more complicated because the discrete octagon
kernel should be modified by a term similar to the last term in
\re{kernelevenbr}.  } Thus we have, for any $\ell$, two operator
representations of the octagon,
\be\begin{aligned} \la{octxrepall} \IO_{ \ell} &= \<\ell |
\exp\(- {1\over 4\pi^2}
\oint_{\mathcal{C}} {dx\over  x}\oint _{\mathcal{C}^*}
{dy\over   y}  
\psi(x)\slK^{\a\a}(x,y) \psi^{*} (y) \) | \ell\>
\quad \quad (\a = \odd, \even) \end{aligned} \ee
with $\slK^{\a\a}$ given by \re{LAurentexp} for $ \a= \odd $ and by
\re{Kernelxeven} for $\a=\even$.  The expectation values \re{octxrepa}
and \re{octxrepabis} are evaluated by the series
 \be\begin{aligned} \la{expandint} \IO_{\ell } &= \sum_{N=0}^\infty
 {(-1)^N\over N!} {1 \over ( 2\pi  )^{ 2N}} \prod_{k=1}^N \oint\limits
 _{\cal{C}}{d x_k\over x_k} \oint\limits _{\cal{C}^*} {d y_k\over y_k}
 \ \({y_j\over x_j}\)^\ell\slK^{\a\a} (x_j, y_j) \ \det_{jk} { x_j \over
 x_j-y_k}
 \end{aligned}   
\ee
 The series \re{expandint} is the expansion of the Fredholm
 determinant of the operator $\slK^{\odd\odd} $ acting in the space of
 the functions, odd for $\a=\odd$ and even for $\a=\even$, which are
 analytic inside the unit circle,

  \be \la{Octg-xspFred} 
  \IO_{ 2m} = \mathrm{Det}\(1-\slK^{\a\a} \), \qquad
  [\slK^{\a\a} f](x ) =\oint {dy\over 2\pi i y}\slK^{\a\a} (x, y)f(y).
   \ee

 The $x$-representation \re{octxrepall}  simplifies a lot in the strong coupling limit, where the leading order  of  the strong
 coupling expansion  can be evaluated using the clustering method \cite{Bargheer:2019exp}, but it is not known how to 
 obtain the  subleading orders.  In this respect  the $\t$-representation is
 more efficient because it  allowed the authers of   \cite{Belitsky:2020qrm,Belitsky:2020qir}  to obtain the  whole strong coupling expansion.   
 On the other hand,   the $x$-representation  allows one to find the 
 strong coupling limit in the case of   more general local weights.

\section{Conclusion}
\la{sect:conclusions}

In this paper we completed the study of the octagon form factor
started in \cite{Kostov:2019stn,Kostov:2019auq} in the following two
aspects.  First, we gave a precise Fock space description of the
fermionic representation outlined there, in which the length of the
bridge determines the level of the Dirac sea.  Second, we found
explicitly the similarity transformation conjectured there, which
leads, by simplifying the octagon kernel, to the determinant formula
for the octagon.  Such similarity transformation is not unique and, as
we already mentioned, another solution has been found independently by
Belitsky and Korchemsky \cite{ Belitsky:2020qir}.  The interpretation
we found for the bridge length allowed us to express the ratio of two
octagons with different bridges as a determinant of finite size
involving the resolvent of the octagon kernel.

The free fermions proposed here as a device to handle the diagonal
symmetric part of the weights of the virtual particles might be useful
for studying other observables which can take the form
\re{octxrepall}, that is, a vacuum expectation value of an element of
$GL(\infty)$. Such objects can be deformed by an infinite set of
commuting flows associated with the modes of the fermion current,
following the recipe of \cite{JimboMiwa-tau}, turning them into
$\tau$-functions of the Toda lattice hierarchy.  Half of these flows
can be associated with the conserved charges in the spin chain
description of $\CN=4$ SYM.

In the case of the octagon, the kinematical parameters $\xi$ and
$\phi$ can be associated, at least in the light-like limit, with the
``times'' coupled to the modes $J_{\pm 1}$ of the fermion current.
Indeed, it was shown in \cite{Belitsky:2020qir} that certain scaling
in the light-like limit the octagon satisfies the radial 2D Toda
lattice equation.  Slightly generalising, we can consider the scaling
limit
 \be \la{newscaling} \phi = i\pi + i {s\over g} , \ \xi = {\s\over g},
 \quad g\to 0 \qquad (s>\s) \ee
 where their solution takes the form
 \be \la{detlcgen} \IO_\ell = e^{- \hat s^2 } \ \det\[ I_{j-k}(2\hat
 s)\]_{j,k=1,..., \ell}, \ee
 where $I_n$ are modified Bessel functions and   $\hat s=\sqrt{ s^2-\s^2}$.
 The octagon in this limit    satisfies  the full 2D Toda lattice equation
with time variable $s$ and space variable $\s$,
\be\la{vraitoda} {1\over 4}\(\p_s^2 - \p_\s^2\) \Phi _\ell +
e^{\Phi_{\ell+1}-\Phi_\ell}- e^{\Phi_\ell-\Phi_{\ell-1}} =0\, , \ \ \
\Phi_\ell= \log {\IO_{\ell+1}\over \IO_{\ell }}\,,  \ \  \ell\ge 1 \,. \ee
A clue about the general validity of \re{vraitoda} would be a direct
proof using the fermion representation, which is still to be found.
     

Finally, let us mention two intriguing recent observations which
suggest to look for a unified formalism working both for the BMN and
the GKP vacua.  First, in the null square limit, the authors of
\cite{Belitsky:2019fan} noticed that the anomalous dimension
characterising the light-like octagon has an alternative
representation similar to the cusp anomalous dimension.  By still
unclear reasons, the light-like limit of the octagon can also be
obtained by choosing as a weight function $\chi(\t) = 2/(e^{\t/g}-1)$.
Second, it was shown in \cite{Basso:2020xts} that the six-gluon
amplitude in certain kinematical limit can be expressed in terms of
the so called { tilted} BES kernel which becomes the BES kernel or the
(light-like) octagon kernel for particular values of the tilting
angle.  The tilted kernel by angle $\a$ is obtained by replacing in
\re{Kmndef} $ i \to i \, e^{-i \a} $ which leads to \re{improvedkernel}
with $\sin\[ (m-n) \pi /2\]$ replaced by $ \sin\[ (m-n) (\pi /2 -\a)\]
$.  The fermionic representation and the subsequent analysis (except
for section \ref{section:dets} which is relevant only to the case
$\a=0$) can be generalised to a generic angle $\a$.  In particular, we
checked that the flow equation obtained in appendix \ref{Appendix:B}
holds for the tilted kernel as well.

\acknowledgments

We thank A. Belitsky and G. Korchemsky for useful discussions and for
sharing their unpublished notes and D. Serban for critical remarks on
the manuscript. V.B.P.  acknowledges the support of 
  the Bulgarian NSF grant DN 18/1.

 \appendix

  \section{Proof of the linear relation \re{Nrelmn}
   between the original and the simplified kernels} \la{Appendix:A}

 We will show that the linear relation \re{transP} holding at the
 level of the integrands \re{originalkernelbis}, \re{improvedkernel},
 leads to the relation \re{Nrelmn} for the integral kernels.  Consider
 the bilinear of Bessel functions
 \be \la{def IJ} \JJ_{m,n}(\t) = i^{m-n-1}
  J_{m}(2\t) J_{n}(2\t) \,.  
  \ee
  With this normalization the functional relation \re{recur0} for the
  Bessel function is rewritten with the help of the matrix $\C$ in
  \re{defCCC} and $\MM_{m,n} = {1\over m}\d_{m, n}$ as \be i { \JJ
  _{m,n}(\t)\over \t}=(\MM \C )_{m,m'} \JJ _{m',n}(\t)= \JJ _{m,
  n'}(\t) (\C \MM )_{n',n}\,, \ \ m,n\ne 0 \,.  \ee Repeated for an
  arbitrary power of $\tau$ this gives, in matrix notations,
\be
 \la{def IJb}
\begin{aligned}
 & \({i\over\t}\) ^j { \JJ (\t) }= (\MM \C )^j \JJ (\t) \,,
\end{aligned}
\ee
where the explicit expressions for the matrix powers of $\MM\C$ are
\be
\la{basicini}
\begin{aligned}
 [ (\MM  \C )^{2s}]_{m,m' }&= \ \sum_{r=-s}^s{2s \choose s- |r|}
 {\Gamma(m-s+r) (m+2r)\over \Gamma(m+s+r+1)} (-1)^{r-s} \delta_{m
 +2r,m' }\,, \\
 [ (\MM  \C )^{2s+1}]_{m,m' }&= \ \sum_{r=-s}^s {2s \choose s - |r|}
 {\Gamma(m-s+r) \over \Gamma(m+s+r+1)}(-1)^{r-s} (\d_{m+ 2r+1,m'
 }-\d_{m +2r-1,m' }).
\end{aligned}
\ee
 The expression \re{basicini} is defined for positive integer $m$ with
 the power $j$ of the matrix $(\MM \C )$ restricted to \be \la{ineq}
 j \le m \,.
\ee
The second $m'$ index in \re{basicini} runs between $m-j$ and $m+j$ (mod
$2$)\,.  For the even power $j=2s$, the formula \re{basicini} has sense
for $j=0$, reproducing the identity $[(\MM 
\C )^{0}]_{m,m'}=\d_{m,m'}$.  Furthermore in this case the formula
extends for $m=0$, taking into account \re{ineq}, i.e., $ [(\MM 
\C )^{2s}]_{0,m'} = \d_{s,0} \d_{0,m'}$.  The powers of $\C\MM$ are
obtained by transposition,
  \be \la{transp} [(\C \MM )^j ]_{n' ,n}& =(-1)^j [( \MM \C )^j
  ]_{n,n' } .  \ee Next we observe that for odd values $j+m - n$, eq.
  \re {def IJb} turns into \be \la{def IJbb}
\begin{aligned}
 &\( {i \over \t}\)^j { \JJ _{m,n}(\t)}=[(\MM \C )^j] _{m,m'}\,
 \CKI(\t)]_{m',n}\, \ \ (j+m= n-1 \ {\rm (mod \ 2) })\,,
\end{aligned}
\ee
where \ $\CKI(\t)_{m,n}= {1-(-1)^{m-n}\over 2} \JJ _{m,n}$
\re{improvedkernel}\,.
Inserting \re{def IJbb} in \re{transP} we obtain after integration
\re{Nrelmn}
\be\la{Nrel}
\begin{aligned}
 \K _{m,n} & = \sum_{ ^{j=0,..., m-
n- 1}_{  j+m+n =\mathrm{odd} }}    {(i
 \s)^j\over j!} A_{m-n}^{(j)} [(\MM \C )^j \KI]_{m ,n}\qquad ( m > n)
 \\
& = \sum_{ ^{j=0,...,  n-m
 - 1}_{  j+m+n =\mathrm{odd} }}  {(-i \s)^j\over j!}
A_{n-m}^{(j)} [\KI (\C \MM )^j]_{m,n}
\qquad (n > m  ) 
  \end{aligned}
 \ee
 The inequality \re{ineq} is fulfilled in \re{Nrel}.  In our problem
 the indices of the Bessel functions $J_m $ take values $m\ge \ell$.
 This implies that the power $j$ of the matrices $(\MM _{ \gell} \C
 _{ \gell} )$ (cf.  \re{defM}) is restricted to
\be
\la{ineql} 
\begin{aligned}
&j \le m-\ell\,,  
\end{aligned}
\ee 
$ [(\MM _{ \gell}  \C _{ \gell} )^{2s}]_{\ell,m'} = \d_{s,0} \d_{\ell,m'}$.
The relation \re{Nrel} then holds with $\MM, \C $ replaced by $ \MM _{ \gell} , \C
_{ \gell}$.  
The restriction of the indices of the
kernel $\K_{m,n}$ given by \re{Nrel} to $m,n\ge \ell $ projects,
taking into account the upper bound \re{ineql}, the second index of
$[(\MM\C)^j]_{m,m'}$ in the rhs to $m'\ge \ell$.

 \section{Flow equation  }
 \la{Appendix:B} 

   The matrix elements  $\K_{m,n}$
   satisfy the following differential equation and its conjugate,   
\be\la{fleqgen}
\begin{aligned}
  m\p_\s \K_{m,n}
  -{i\s }\p_\s \(  \K_{m+1, n} +   \K_{m-1,n}\) 
  + i  {(m-n) }    \(  \K_{m+1,n}- \K_{m-1,n}\)&=0\, ,
  \end{aligned}
\ee
for any $m,n\ge 0$.  Here the weight function $\chi$ is treated as a
functional parameter and the derivative in $\s$ does not act on it.
The equations follow straightforwardly from \re{Nrelmn} using the
relations for the coefficients \be A^{j+1}_{m-n} = - (m-n\mp j)
A^j_{m\pm1 -n}, \quad A^{m-n}_{m+1-n}=0.  \ee Introducing the diagonal
matrix $\NN _{m,n} = {n }\, \d_{m,n}\,, m,n\ge 0$ the equation
\re{fleqgen} and its conjugate can be cast in a matrix form \be
 \begin{aligned}
 (\NN- i \s \SS)\p_\s\K+i [\NN, \C \K] &=0 \\
 \p_\s\K\ (\NN- i \s  \SS) + i  [ \NN ,  \K\C]&=0.
\la{shortflowmat}
\end{aligned}
\ee

 The flow equation \re{shortflowmat} determines the evolution of the
 full kernel $\K$ which characterises the octagon with zero bridge.
 The equation for non-zero bridge is obtained by replacing the kernel
 and the matrices involved with $ \K_{ \gell} \,, \NN_{ \gell}  \,,
 \SS_{ \gell} \,,\C_{ \gell} \,.$

\medskip
 
 \noindent {\it Remark.} After substituting $ \K_{ \gell} \to \UU_\ell
 ^{\mathrm{T}} \KI_{ \gell} \UU_\ell$, the equation \re{fleqgen} turns
 into an equation for the similarity operator $\UU_\ell$.  In general,
 it is not straightforward to integrate it.  Since there is a
 continuum of solutions, one can impose additional conditions on the
 solution.  Belitsky and Korchemsky imposed \cite{ Belitsky:2020qir}
 the condition that the semi-infinite similarity matrix $\OO=
 \{\OO_{i,k}\}_{k,j\ge 0}$ acts trivially on the first two columns,
 \be
 \la{GBcond} \OO_{k, 0} =\d_{k,0}\,, \OO_{ k , 1} =\d_{k,1}\,.
 \ee  In our conventions
 their matrix $\OO$ corresponds to a matrix $ \tilde \UU _\ell$ with
 elements
  \be
 [\tilde  \UU _\ell ]_{k+\ell, j+\ell} = [\OO]_{k,j}.
 \ee
  Under these conditions $[\tilde\UU_\ell ]_{k+\ell, \ell} =\d_{k,0}$
  and $ [\tilde\UU_\ell ]_{k+\ell, 1+\ell} =\d_{k,1}$ one obtains
  \be \la{Nrel0}
  \begin{aligned}
  \K_{m+\ell,\ell}&= [\tilde\UU_\ell ^\mathrm{T}\,  \KI]_{m+\ell,\ell}\
  , \quad \K_{m+\ell,1+\ell}&= [\tilde\UU^\mathrm{T}_\ell\, 
  \KI]_{m+\ell,1+\ell} \qquad (m\ge 0).
  \end{aligned}
 \ee 
 In this way the differential equation \re{fleqgen} reduces to an
 equation for $\tilde\UU_\ell$.  Accordingly the first/second relation
 \re{Nrel0} determines the matrix elements of $ \OO^\mathrm{T} $ with
 odd/even first index directly from \re{Nrelmn}
\be
\la{BG}
\begin{aligned}
 & \OO_{2k+1, n}= \sum_{^{ j =0,..., n-1}_{ j-n = {\rm \ odd}} }  {(-i
 \s)^j\over j!} [(\C \MM )^{j}]_{2k+1+\ell,n+\ell} A_{n}^{(j)} =
 [\UU _\ell]_{2k+1+\ell, n+\ell}
 \\
  & \OO_{2k, n} = \d_{n,0}\d_{k,0}+\sum_{ ^{j =0,..., n-2}_{ j- n =
  {\rm \ even} }} {(-i \s)^j\over j!} [(\C \MM
  )^{j}]_{2k+\ell,n+\ell} A_{|n-1|}^{(j)} \\
  &= \d_{n,0}\d_{k,0}+  [\UU _{\ell+1}]_{2k-1+(\ell+1), n-1+(\ell+1)}\,.
 \end{aligned}
\ee
The last equality relates the matrix elements of $\OO$ to a
different projection of the solution $ [\UU_\ell]_{n',n} $ \re{Nrel1}
to odd $n'-\ell $. 

  \section{Proof of the similarity transformation \re{similaritytrgen}
  - \re{Nrel1}} \la{Appendix:C}

 In this appendix we will omit the index $\ell$ in $\UU_\ell$ and $\MM
 _\ell$ in order to avoid ugly formulas.  We will show that the linear
 transformation \re{Nrelmn} can be written as adjoint action matrix
 relation \re{similaritytrgen}
\be \la{simil2} \K_{m+\ell,n+\ell} =\sum_{^{m' ,n' \ge 0}_{ n' -m' =
\mathrm{odd} }}
\UU^{\mathrm{T}}_{m+\ell,m'+\ell}\KI_{m'+\ell,n'+\ell}\UU_{n'+\ell,n+\ell}\,.
\ee
The matrix elements of $\UU= \UU_\ell$ in \re {Nrel1} read more
explicitly
\be
\la{operatorsY}
\begin{aligned}
 \UU_{2k+1+\ell, n+\ell } &= \sum_{ ^{ j= 0,..., n-1 }_{ j -n =
 \mathrm{odd} }} {(-i \s)^j\over j!} \ A^{(j)}_{n} \ [(\C \MM
 )^j]_{2k +1+\ell,n+\ell}\,\\
   \UU_{2k+\ell, n+\ell }& = \sum_{ ^{ j= 0,..., n }_{ j -n =
   \mathrm{even} }} {(i \s)^j\over j!} \ B^{(j)}_{n} [\(\C \MM
   \)^j]_{2k+\ell,n+\ell}\,.  \\
\end{aligned}
\ee
They satisfy the relations 
\be \la{boundary}
\begin{split}
  \UU_{ k+\ell,\ell}&=\d_{k,0}\,, \quad \UU_{ 2k+1+\ell,1+\ell}
  =\d_{k,0} \,,
    \end{split}
\ee 
which differ from \re{GBcond} since $\UU_{
2k+\ell,1+\ell}=i\s(\d_{k,0}-\d_{k,1})$. 
Note that the zeros of the coefficients $A^{(j)}_n$ and $B^{(j)}_n $
in \re{relB-A} compensate the poles in the expressions \re{basicini} for
the matrix powers $(\C \MM )^j_{n'+\ell,n+\ell} $ and one can write
regularised closed expressions for the operators \re{operatorsY}.  For
example\footnote{ This expression does not change if the upper bound
of the summation is extended to $ s\le n+m+\ell -1$.}
  \be
  \begin{aligned}
&\UU_{2m+\ell, 2n+\ell} = \d_{ m,0}\ \s^{2 n} (2^{2 n-1} (-1)^n\,
(1-\d_{n,0}) +\d_{n,0}){\G(1+\ell)\over \G(2 n +\ell+1)} \\
& + \theta_m \sum_{s=0}^{n} {(2\s)^{2s} (-1)^{m + n} (2m+\ell)\, n\,
\G( n +s) \prod_{k=1}^{m+\ell-1}(n -s + k) \over \G(s- m + n +1)\G(s+m
- n +1) \G(n +m +\ell+s+1) }.
  \end{aligned}
 \ee

We want to prove \re{simil2} with the operators given in
\re{operatorsY}.  The key ingredient of the proof is the intertwining
relation
   \be
 \la{intertwine}
\begin{aligned}
& [(\MM \C )^{j}{ \KI}]_{m+\ell,n+\ell} =[(\MM \C )^{j-p} {
\KI}(\C \MM )^p]_{m+\ell,n+\ell} \ \qquad ( j\le m \,, \ p\le n ).
\end{aligned}
\ee
The latter allows to redistribute the matrix powers in \re{Nrel} on
both sides of $\KI$.  E.g., 
\be \la{Kmn12}
\begin{aligned}
-A^{(1)}_{m-n} [\MM\C\, \KI]_{m,n}&=(m-n) [\MM\C\, \KI]_{m,n}=[m\,
\MM\C\ \KI)- n\, \KI\, \C\MM]_{m,n} \\
\, A^{(2)}_{m-n} [(\MM\C)^2\, \KI]_{m,n} &= [m^2 (\MM\C)^2\, \KI-
2 m(\MM\C) \KI(\C \MM)n+ (n^2\!-\!1) \KI(\C\MM)^2]_{m,n}\,.
\end{aligned}
\ee
In what follows we assume, for the sake of simplicity of the 
presentation, that $\ell=0$.  To illustrate the general procedure,
consider the matrix element $\K_{1,n}$\,.  We can split the
coefficient in  \re{Nrelmn} as 
\be \la{split1} A_{n-1}^{(j)} =\sum_{p=0}^j
   {j \choose p} B_1^{(p)} A_n^{(j-p)}  =j\,
B_1^{(1)}A_{n}^{(j-1)} +(-1)^j B_n^{(j)} A_1^{(0)} \ee
taking into account that $B_1^{(2s+1)}=0 $ for $s  \ge 1$, while the contributions of $B_1^{(2s)}\,, s\ge 0 $ sum to the second term in the rhs. Next we  distribute
accordingly the matrix powers in agreement with the inequality
\re{ineq}
\be \la{newrelred}
\begin{aligned} 
&{\K}_{1,n} = \KI_{1,n'} \sum_{ ^{ j= 0, ..., n-2 }_{j-n=
\mathrm{even} }} {(-i \s)^j\over j!} [(\C \MM )^{j}]_{n',n}
A_{|n-1|}^{(j)}\\
 &= \!\!\!\!\!  \sum_{{}^{ j=0,..., n-2}_{j-n= \mathrm{even} }} {(-i
 \s)^{j}\over j!}\, \Big( j\, B_1^{(1)} [(\MM \C ) \, \KI (\C \MM
 )^{(j-1)}]_{1,n} \, A_n^{(j-1)} +(-1)^j [\KI (\C \MM )^{(j)}]_{1,n}
 B_n^{(j)} \Big) \,.\ \ \
\end{aligned}
\ee
This is almost \re{simil2} in this particular case, with odd
intermediate summation index $n' $ in the first term in the rhs of
\re{newrelred} and $n'$- even in the second.  The two terms reproduce
the corresponding expressions for the operators in \re{operatorsY}
up to the upper bounds.  In fact the upper bound in \re{newrelred}
extends from $n-2$ to $n$.  In the first line this is so due the
vanishing of $A^{(n)}_{n-1}=0$.  Equivalently for $j=n$ the two terms
in \re{split1} and their contributions to \re{newrelred} compensate
each other due to the relation \re{relB-A}.  Hence, taking into
account \re{boundary} which implies  $\UU^\mathrm{T}_{1,2k+1} =\delta_{k,0}$, we reproduce
\re{simil2} for this particular example
\be \la{newrel1}
\begin{aligned} 
&\K_{1,n}= [\UU^\mathrm{T}\KI \UU]_{1,n} = \sum_{n' - {\rm \ odd } }
\UU^\mathrm{T}_{1,m'} \KI _{m', n'} \UU_{n', n} + \sum_{n' - {\rm \
even} } \UU^\mathrm{T}_{1,m'} \KI_{m',n'} \UU_{n', n}\, .
\end{aligned}
\ee 

The generalization of \re{split1} for $m<n $ reads
\be
\la{splittgene}
\begin{aligned}
A_{n-m}^{(j)} =\sum_{^{s=0, ..., m}_{s-m= \mathrm{even}} } {j\choose
s} B_m^{(s)} A_n^{(j-s)}+(-1)^j \sum_{^{s=0, ..., m-1}_{s-m=
\mathrm{odd}} } {j\choose s} A_m^{(s)} B_n^{(j-s)}\,.
\end{aligned}
\ee
For general $\K_{m,n}$ one proceeds as in the example $\K_{1,n}$
considered above, distributing accordingly the matrix powers to the
left and right of $\KI$.  The last step is to ensure the upper bounds
as in \re{operatorsY}.  E.g., for odd $m$ the upper bound in the
initial expression \re{Nrel} can be lifted to $n$ adding $m+1$ terms
without violating \re{ineq} exploiting the vanishing of the
coefficients $A^{(j)}_{n-m}$ for $j\ge n-m+1 +2k\,, k\in \IZ_+$. 
 Moreover,
the upper bound can be extended further to $n+m-1$ moving to the left
the additional matrix powers of $(\C \MM )$, so that to comply with
the inequalities \re{ineq}.  In the process, some zeros may appear
also in each of the two terms in \re{splittgene}.  Altogether this
ensures the correct upper bounds of the operators $\UU $ in the
transformed expresssion obtained using \re{splittgene}.

\section{Proof of  the exponential representation for $\ell=0$ }
\la{Appendix:D}

Here we show that the matrix $\UU $ defined in \re{Nrel1} 
 is given by the series  \re{Fullexponentialsolution} for
$\ell=0$.  The matrix $\UU $ and its transposed $ \UU ^{\mathrm{T}}$
factorise as
\begin{align}
\la{blockVV} \UU &=\hat\UU\PP , \quad \hat \UU = \(\PP _\even\
e^{-{1\over 2} \s^2 \C\MM\SS} + \PP _\odd \ e^{-{1\over 2} \s^2
\SS\C\MM} \)e^{i\s\C}= \sum _{j\ge 0} [\hat \UU]_j \s^j \ ; \\
 \UU
^{\mathrm{T}}&=\PP\ \hat\UU^{\mathrm{T}},
\quad
\hat\UU^{\mathrm{T}}=
e^{-i\s\C}\( e^{{1\over 2} \s^2 \SS\MM\C}
\PP _\even\ +\ e^{{1\over 2} \s^2 \MM\C\SS} \ \PP ^\odd \)
=
\sum _{j\ge 0} [ \hat \UU ^{\mathrm{T}}]_j\s^j \ , 
\la{expUT}  \end{align}  
where $\PP$ is the projector restricting the power $j$ of $\s$ of the
matrix element $[\hat\UU]_{n',n}$ to $n$ or $n-1$ as in
\re{Fullexponentialsolution}.

 We will give the idea of the proof for the transposed matrix
 $\hat\UU^{\mathrm{T}}$, restricting ourselves to the piece acting in
 the even sector.  The coefficients $\XX^{(j)} = [\hat
 \UU^{\mathrm{T}}]_j \PP_\even$ in the expansion of the first term in
 \re{expUT} read
\be \la{beta1} j!  \XX^{(j)} = j!\sum_{k=0}^{[{j\over 2}]} \b_k^j \C
^{j-2k} (\SS \MM \C )^k\,, \ \ \ \b_k^j = {(-1)^k \over 2^k (j-2k)!
k! }\,.  \ee
On the other hand, the piece of the transposed matrix \re{Nrel1}
restricted to the even sector, $\UU^{\mathrm{T}}_\even=
\UU^{\mathrm{T}}\PP_\even$~, is expanded as
\be [\UU^{\mathrm{T}}_\even]_{m,m'}= \sum_{^{j =0,...,
m}_{j-m=\mathrm{even}}} {(-i\s)^j\over j!}B^{(j)}_m
[(\MM\C)^j]_{m,m'}\,, \qquad m'- {\rm \ even} .
\ee
We have to show that
\be \la{expB} B_{m}^{(j)} [(\MM \C )^{j}]_{m,m'} =j!  \XX^{(j)}
_{m,m'} \, .  \ee
 Let us see how this works with the lowest coefficients $j=1,2$.  By
 the expression \re{ABcoef} for the coefficients $B^{(j)}_m $, for $
 j=1, 2$ we have
 \be \la{exam12}
\begin{aligned}
&B^{(1)}_m  (\MM\C)_{m,m'}= m   (\MM\C)_{m,m'}=\C_{m,m'}\,, \\
& B^{(2)}_m [(\MM\C)^2]_{m,m'}= m^2
[(\MM\C)^2]_{m,m'}=m\sum_{\pm}\pm [\MM\C]_{m\pm 1,m'}\\
&=\sum_{\pm} \pm \((m\pm1) \mp 1\)[\MM\C]_{m\pm 1,m'}=[\C^2
-\SS\MM\C]_{m,m'}\,.
\end{aligned}
\ee
We see that the computation for $j=2$ in the last line of \re{exam12}
is reduced to that for $j=1$.  The general proof of \re{expB} can be
done by induction.  For that we will use the following recursive
formula for $B^{(j)}_m $
\be
\la{recurB}
\begin{aligned}
 B_m^{(j+1)}&= (-1)^j m A_{m\pm 1\mp 1} ^{(j)} = m \(B_{m\pm 1}^{(j)}
\pm \sum_{^{p =1,..., j }_{ p= \mathrm{odd} } } {j\choose p} B_{m\pm
1}^{(j-p)} A_{1}
^{(p)} \) \\
&= m \(B_{m\pm 1}^{(j)} \mp \sum_{^{p =1,..., j }_{ p= \mathrm{odd} }
} {j\choose p} \prod_{s=0}^{p-1\over 2} (1- (2s)^2)\, B_{m\pm
1}^{(j-p)}\) 
\end{aligned}
\ee 
derived from the expansion
\be
\la{splitt}
\begin{aligned} 
   & A^{(j)} _{m-n} =(-1)^j \sum_{p=0}^j {j \choose p} B_m^{(p)}
   A_n^{(j-p)} \ \ \qquad (m>n)
\end{aligned}
\ee
and we have taken into account that $ A_{-1} ^{(p)} = - A_{1} ^{(p)} $ and
 $A_{1} ^{(2p)}=\d_{p,0} $. 
 From \re{recurB} we obtain a recursive
 formula for the corresponding matrices,
  \be \la{recurs}
\begin{aligned}
&
B_m^{(j+1)}((\MM \C )^{j+1})_{m,n}= \sum_{\pm} \pm (B_{m\pm
1}^{(j)} [(\MM \C )^{j}]_{m\pm 1,n} \\
& - \sum_{^{p =1,..., j }_{ p= \mathrm{odd} } } {j\choose p}
\prod_{s=0}^{p-1\over 2} (1- (2s)^2)\, \sum_{\pm} (B_{m\pm 1}^{(j-p)}
[(\MM \C )^{j-p} (\MM \C ) ^{p}]_{m\pm 1,n} \\
\end{aligned}
\ee 
If we assume the relation \re{expB}, then \re{recurs} implies that the
coefficients $\XX^{(j)}$ satisfy the recurrence relation \re{recurs}
which takes the form
\be 
\la{recursi}
\begin{aligned}
(j+1)!\ \XX^{(j+1)}&= j!\C \XX^{(j)} - j (j-1)!  \ \XX^{(j-1)} (\SS \MM
\C )^{1} +\CA ,
\end{aligned}
\ee
where we wrote explicitly only the first two terms.  If we can prove
independently that \re{recursi} is satisfied, this would imply
\re{expB}.  To do that, let us first notice that the lhs of
\re{recursi} equals the sum of the first two terms in the rhs.  This
follows from the explicit form of $\XX^{(j)}$, eq.  \re{beta1}.
Therefore to prove \re{expB} it is sufficient to show that $\CA=0$.
One can check, after tedious algebra, that this is indeed the case.
Let us only write down a basic commutator used:
 \be
\begin{aligned}
[\SS ,(\SS \MM \C )^k] &= -\sum_{r=0}^{k-1} a_r^k (\SS \MM \C
)^{k-r} (\MM \C )^{2r+1}\,,
\\
a_r^k &= \prod_{s=1}^r (2s-1) {k\choose r+1}\,, \ a_0^k=k .
\end{aligned}
\ee
 One of
 the nice features of the exponential form \re{blockVV} is that it
 renders the symplectic property $\C = \UU \C \UU ^{\mathrm{T}}$
 almost obvious,
\be\begin{aligned} \UU\C \UU^\mathrm{T} & = \(\PP _\even\ e^{-{1\over 2} \s^2
\C\MM\SS} + \PP _\odd \ e^{-{1\over 2} \s^2 \C\SS\MM} \) \PP\C
\PP\( e^{{1\over 2} \s^2 \SS\MM\C} \PP _\even\ +\ e^{{1\over 2} \s^2
\MM\C\SS} \ \PP _\odd \) \\
& = \PP _\even\ e^{-{1\over 2} \s^2 \C\MM\SS} \ \PP \ e^{ {1\over 2}
\s^2 \C\MM\SS} \C + \PP _\odd \ e^{-{1\over 2} \s^2 \C\SS\MM}\ \PP
\ e^{{1\over 2} \s^2 \C\SS\MM}\C \\
&=(\PP_\even+\PP_\odd)\PP \C \ =\C.\ \ 
 \end{aligned}
  \ee
%

 
 \providecommand{\href}[2]{#2}\begingroup\raggedright\endgroup

\end{document}